\documentclass[lettersize,journal]{IEEEtran}

\usepackage{verbatim}
\usepackage{enumerate}
\usepackage{amssymb}
\usepackage{amsthm}
\usepackage{multirow}
\usepackage{xcolor}
\usepackage{cite}
\usepackage{array}

\usepackage{graphicx}
\usepackage{epstopdf}

\usepackage{amsmath,amsfonts}
\usepackage{algorithmic}
\usepackage{algorithm}

\usepackage{textcomp}
\usepackage{stfloats}
\usepackage{url}

\newtheorem{construction}{Construction}
\newtheorem{theorem}{Theorem}

\newtheorem{lemma}{Lemma}
\newtheorem{definition}{Definition}
\newtheorem{remark}{Remark}

\begin{document}

\title{Variant Codes Based on A Special Polynomial Ring and Their Fast Computations}

\author{Leilei~Yu,~Yunghsiang S. Han,~\IEEEmembership{Fellow,~IEEE}, Jiasheng Yuan, and Zhongpei Zhang
	\thanks{L. Yu, Y. S. Han, J. Yuan and Z. Zhang are with the Shenzhen Institute for Advanced Study, University of Electronic Science and Technology of China, Shenzhen, China (e-mail: yuleilei@uestc.edu.cn, yunghsiangh@gmail.com, 202312281024@std.uestc.edu.cn, Zhangzp@uestc.edu.cn).
	}
}

\markboth{Journal of \LaTeX\ Class Files,~Vol.~x, No.~x, August~xxxx}%
{Shell \MakeLowercase{\textit{et al.}}: A Sample Article Using IEEEtran.cls for IEEE Journals}


\maketitle

\begin{abstract}

Binary array codes are widely used in storage systems to prevent data loss, such as the Redundant Array of Independent Disks~(RAID). Most designs for such codes, such as Blaum-Roth~(BR) codes and Independent-Parity~(IP) codes, are carried out on the polynomial ring $\mathbb{F}_2[x]/\langle \sum_{i=0}^{p-1}x^i \rangle$, where $\mathbb{F}_2$ is a binary field, and $p$ is a prime number.
In this paper, we consider the polynomial ring $\mathbb{F}_2[x]/\langle \sum_{i=0}^{p-1}x^{i\tau}\rangle$, where $p>1$ is an odd number and $\tau\geq 1$ is any power of two, and explore variant codes from codes over this polynomial ring.
Particularly, the variant codes are derived by mapping parity-check matrices over the polynomial ring to binary parity-check matrices.

Specifically, we first propose two classes of variant codes, termed V-ETBR and V-ESIP codes.
To make these variant codes binary maximum distance separable~(MDS) array codes that achieve optimal storage efficiency, this paper then derives the connections between them and their counterparts over polynomial rings. 
These connections are general, making it easy to construct variant MDS array codes from various forms of matrices over polynomial rings.
Subsequently, some instances are explicitly constructed based on Cauchy and Vandermonde matrices.
In the proposed constructions, both V-ETBR and V-ESIP MDS array codes can have any number of parity columns and have the total number of data columns of exponential order with respect to $p$.
In contrast, previous binary MDS array codes only have a total number of data columns of linear order with respect to $p$. 
This makes the codes proposed in this paper more suitable for application to large-scale storage systems.
In terms of computation, two fast syndrome computations are proposed for the Vandermonde-based V-ETBR and V-ESIP MDS array codes, both meeting the lowest known asymptotic complexity among MDS codes.
Due to the fact that all variant codes are constructed from parity-check matrices over simple binary fields instead of polynomial rings, they are attractive in practice.

\end{abstract}

\begin{IEEEkeywords}
	Storage systems, binary array code, binary parity-check matrix, syndrome computation.
\end{IEEEkeywords}

\section{Introduction}\label{sec:1}
\IEEEPARstart{M}odern distributed storage systems require data redundancy to maintain data reliability and durability in the presence of unpredictable failures.
Replications and erasure codes are two typical redundancy mechanisms~\cite{yu2023reed,hou2022generalization}.
Compared to the former, erasure codes only need less data redundancy to attain the same level of data protection~\cite{cook2014compare}.
One well-known class of erasure codes is \textit{binary array codes}~\cite{blaum1993new,blaum1996mds,blaum2006family,hou2016mds}.
Their coding procedures involve only XOR~(exclusive OR) and cyclic shift operations,
which enables simple and efficient implementations in both software and hardware~\cite{lv2022new}.
This paper focuses on such codes.

Binary array codes have been widely used in storage systems, such as RAID~(Redundant Array of Independent Disks)~\cite{patterson1989introduction}.
With the development of distributed storage systems in recent years, they have also been used as the basis for developing other erasure codes, such as locally repairable codes~\cite{lv2022new,blaum2019array,hou2022generalization,blaum2013partial} and regenerating codes~\cite{shum2014regenerating,ye2016explicit,hou2022towards}.
For an $\ell\times (k+r)$ binary array code, any codeword can be viewed as an $\ell\times (k+r)$ array of bits, where $k$ columns store all information bits to form $k$ information columns, and the remaining columns store all the parity bits encoded from information bits to form $r$ parity columns.
The row size $\ell$ generally depends on the code construction.
In coding theory, maximum distance separable~(MDS) codes reach optimal storage efficiency~\cite{shen2014hv}, and each of their codewords consists of information and parity symbols, such that any subset of symbols in the codeword with the same number as information symbols can recover the entire codeword.
\textit{Binary MDS array codes} have the same property by treating each column as a symbol. More precisely, for an $\ell\times (k+r)$ binary MDS array code, any $k$ out of $k+r$ columns suffice to decode~(reconstruct) all columns. 
Some well-known examples of binary array codes are EVENODD~\cite{blaum1995evenodd},
row-diagonal parity~(RDP)~\cite{corbett2004row}, STAR~\cite{huang2008star}, and triple-fault-tolerance codes~\cite{hou2017triple}.
These codes are all binary MDS array codes for the case of two or three parity columns.
Examples of binary array codes with more parity columns are Blaum-Roth~(BR)~\cite{blaum1993new}, Independent-Parity~(IP)~\cite{blaum1996mds}, generalized RDP codes~\cite{blaum2006family}, and the codes in~\cite{hou2014new}.
Although they are not always binary MDS array codes, the conditions that render them such codes can be found in the corresponding literature.

The new binary array codes proposed in this paper target an arbitrary number of parity columns, and their constructions are closely related to the BR, IP, and generalized RDP codes mentioned above. 
Specifically, BR and IP codes are both constructed by parity-check matrices over the polynomial ring ${\mathbb{F}_2[x]}/{\langle \sum_{i=0}^{p-1}x^i \rangle}$, where $\mathbb{F}_2$ denotes a binary field and $p$ is a prime number~\cite{blaum1993new,blaum1996mds}. 
Generalized RDP codes can be regarded as a variant of shortened IP codes~\cite{blaum2006family}, and they possess lower computational complexity~\cite{huang2016improved}.
In this paper, we reformulate the generalized RDP codes, and then one can intuitively understand the essence of the generalized RDP codes being more computationally superior.
Briefly, when computing syndromes,
the codes over $  {\mathbb{F}_2[x]}/{\langle \sum_{i=0}^{p-1}x^i \rangle}  $ are first calculated in an auxiliary polynomial ring $ {\mathbb{F}_2[x]}/{\langle x^p+1\rangle}  $, where multiplying $x$ only requires performing a simple cyclic shift operation. Then all results are returned to the original ring~\cite{blaum1993new,blaum1996mds}.
As a variant, the generalized RDP codes have a similar process to the shortened IP codes in computing syndromes, with the only difference being that they do not process the extra bits of the auxiliary polynomial ring compared to the original ring.
Thus, the generalized RDP codes eliminate two operations in the shortened IP codes when computing syndromes. One is the processing for one fixed bit in each symbol over the auxiliary ring, and the other is the modulo operation for returning to the original ring.
A binary parity-check matrix for the generalized RDP codes is explicitly provided in this paper (Please refer to \eqref{eq:15}).

In fact, this paper generalizes the above variant technique so that new codes based on binary parity-check matrices can be easily obtained from codes over the polynomial ring $ {\mathbb{F}_2[x]}/{\langle \sum_{i=0}^{p-1}x^{i\tau}\rangle}  $, where $p$ is an odd number and $\tau$ is any power of two.
In our setup, the parity-check matrices of codes over the polynomial ring can be determined not only by the Vandermonde matrices containing only monomials~(e.g. BR, IP codes) but also by matrices with more forms (e.g. Cauchy matrices, etc.) and wider parameter ranges.	
In this paper, two classes of codes defined in ${\mathbb{F}_2[x]}/{\langle \sum_{i=0}^{p-1}x^{i\tau}\rangle} $ are referred to as ETBR and ESIP codes, which can be regarded as extensions of BR and shortened IP codes, respectively.
Correspondingly, the variants of ETBR and ESIP codes are referred to as V-ETBR and V-ESIP codes, respectively.
The main contributions of this paper are enumerated as follows:

\begin{enumerate}
	\item 
	
	This paper proposes two new classes of binary array codes~(i.e., V-ETBR and V-ESIP codes), which are both based on binary parity-check matrices~({see Sec.~\ref{sec:3}}). 
	We show that the well-known generalized RDP codes are a special case of the V-ESIP codes.
	
	\item This paper presents the conditions for the new codes to be binary MDS array codes by exploring the connections between them and their counterparts over the polynomial ring~({see Sec.~\ref{sec:iv}}).
	In particular, these connections are built on the foundation that all parity-check matrices have a sufficiently flexible form.
	This provides convenience for constructing V-ETBR/V-ESIP MDS array codes with various forms.
	
	\item Based on Vandermonde and Cauchy matrices, this paper explicitly provides the constructions for the V-ETBR and V-ESIP MDS array codes, both with any number of parity columns $r$~({see Sec.~\ref{sec:3.b}}).
	Compared to previous binary MDS array codes over the polynomial ring, the constructed codes have significantly more data columns for a given design parameter $p$, as well as a more flexible row size $\ell$.
	
	\item This paper also proposes two fast syndrome computations, which respectively correspond to the V-ETBR MDS array codes with any $r\geq 2$~({see Sec.~\ref{sec:v.a}}) and the V-ESIP MDS array codes with $r=4$~({see Sec.~\ref{sec:v.b}}).
	Both of them meet the lowest known asymptotic computational complexity among MDS codes~\cite{yu2023reed}, i.e., each data bit requires $\lfloor\lg r\rfloor+1$ XORs as the total number of data columns approaches infinity.
	
\end{enumerate}

%
	In this paper, the proposed fast syndrome computations can be seen as an extension of the syndrome computation in Reed-Solomon~(RS) codes over finite fields \cite{yu2023reed} to the variant codes.
	In \cite{yu2023reed}, the computation involved in RS codes can generate a large amount of intermediate data through the Reed-Muller~(RM) transform to reduce the total number of operations.
	Some variant codes constructed in this paper are based on Vandermonde matrices (over polynomial rings) with a similar structure as in \cite{yu2023reed}, and the fast computation in RS codes is compatible with these constructed variant codes.
In this paper, the fast computations proposed for variant codes can be easily adjusted to be suitable for the corresponding codes over the polynomial ring. 
To avoid tediousness, we will not repeat the presentation.
Note that the variant codes are based on binary parity-check matrices, leading to easy implementation through the use of existing open-source libraries for matrix operations over $\mathbb{F}_2$, such as M4RI~\cite{albrecht2012m4ri}.
This means that engineers can use them without needing to have much knowledge of algebra.
At the end of this paper, we also compared the specific number of XORs required for encoding and decoding of the variant codes with other alternative binary MDS array codes, i.e., Circulant Cauchy code~\cite{schindelhauer2013maximum}, Rabin-like code~\cite{hou2017new}, and BR code~\cite{blaum1993new,subedi2013comprehensive}.
When the total number of data columns is 251, and the number of parity columns ranges from 4 to 7, the average encoding/decoding improvements of variant codes compared to them are 69\%/69\%, 63\%/61\%, and 26\%/22\%, respectively.
Since the variant codes are based on simple binary parity-check matrices, there is still a great potential to further improve computational efficiency by using scheduling algorithms for binary matrix multiplication, such as~\cite{plank2012heuristics,huang2007optimizing}, etc.

Recently, \cite{lv2022new,lv2023binary}, and \cite{lv2023new} proposed some new binary MDS array codes.
Their idea is to construct binary parity-check matrices by truncating circulant matrices of elements over polynomial rings. 
The resulting binary MDS array codes are essentially V-ETBR/V-ESIP codes, and this paper can be seen as a generalization of their works.
This generalization extends parity-check matrices restricted to Vandermonde forms to having arbitrary matrix forms, as well as extends the Vandermonde-based syndrome computation in their works, which is only applicable to $2\leq r\leq3$, to supporting arbitrary $r\geq 2$.
Furthermore, one of the main contributions of this paper is to propose the intrinsic connections between codes over the polynomial ring and V-ETBR/V-ESIP codes.
This was not considered in the previous work.
Particularly, these connections provide a powerful tool for constructing binary MDS array codes over binary fields.
%
The detailed differences between the previous work and this paper are enumerated as follows:

\begin{enumerate}	
	\item This paper clearly reveals the relationship between V-ETBR/V-ESIP codes and the well-known generalized RDP codes, as the former is a generalization of the variant technique implied by the latter. This was not pointed out in the previous work.
	\item In the previous work, the V-ETBR/V-ESIP codes consider only binary parity-check matrices determined by Vandermonde matrices.
	In contrast, the matrices used in this paper have a more flexible form, of which the Vandermonde matrix is just a special instance. This can facilitate the construction of more variant codes.
	\item The previous work focuses only on V-ETBR/V-ESIP codes without discussing their connections with the corresponding codes over polynomial rings. In this paper, we consider these connections and show that, based on them, new MDS codes over polynomial rings can be directly obtained as by-products.	
	\item In terms of construction, all MDS array codes proposed in the previous work and this paper can have a total number of data columns far exceeding the design parameter $p$. However, the feasible number of parity columns for the V-ESIP MDS array codes in the previous work is three, while that in this paper is any size.
	\item In terms of computation, fast syndrome computation in the previous work is for $2\leq r\leq 3$, whereas that proposed in this paper is for arbitrary $r\geq 2$. The former is a special case of the latter.
\end{enumerate}

The remainder of this paper is organized as follows. Section~\ref{sec:2} introduces all necessary preliminaries, including some existing well-known binary array codes and important notations. Section~\ref{sec:3} provides the specific definitions of ESIP/ESIP and V-ETBR/V-ESIP codes.
{By exploring the general connections between V-ETBR/V-ESIP codes and their counterparts over polynomial rings~(i.e., ESIP/ESIP codes), Section~\ref{sec:iv} proposes the conditions that make variant codes binary MDS array codes.
In Section~\ref{sec:3.b}, some explicit constructions for V-ETBR and V-ESIP MDS array codes are proposed, along with their fast syndrome computations.}
Section~\ref{sec:5} concludes this paper.

\section{Preliminaries}\label{sec:2}
This section describes some existing well-known classes of array codes, i.e., BR codes~\cite{blaum1993new}, IP codes~\cite{blaum1996mds}, and generalized RDP codes~\cite{blaum2006family}.
To begin with, let
\begin{equation}\label{eq:11}
	\mathbb{R}_{p,\tau}:=\frac{\mathbb{F}_2[x]}{\langle f_{p,\tau}(x)\rangle}
\end{equation} 
denote a binary polynomial ring,
where 
\begin{equation}\label{eq:02}
	f_{p,\tau}(x)=1+x^{\tau}+\cdots +x^{(p-1)\tau}
\end{equation}
with two positive integers $p,\tau$.
The identity that $x^{p\tau}+1= (x^\tau+1)\cdot f_{p,\tau}(x)$ leads to operations in $\mathbb{R}_{p,\tau}$ that can be performed first in polynomial ring 
\begin{equation}\label{eq:R}
	\mathbb{R}:=\frac{\mathbb{F}_2[x]}{\langle x^{p\tau}+1 \rangle} \;,
\end{equation}
and then, all results should be reduced modulo $f_{p,\tau}(x)$.
Since multiplying by $x$ in $\mathbb{R}$ is equivalent to performing a one-bit cyclic shift on a vector with $p\tau$ bits, the above realization for the operations in $\mathbb{R}_{p,\tau}$ is simple and efficient~\cite{blaum1993new,blaum1996mds}.

\subsection{BR codes}\label{sec:2a}
BR codes are constructed in polynomial ring $\mathbb{R}_{p,1}$~\cite{blaum1993new}, where $p$ is a prime number.
Given the value of $p$, the BR$(p,r<p)$ is defined as the set of $(p-1)\times p$ arrays (denoted by $[x_{i,j}]$, where $x_{i,j}\in \{0,1\}$, the first $p-r$ data columns are information columns and others are parity columns).
For $\ell=0,1,...,p-1$, the $\ell$-th column of a $(p-1)\times p$ array can be viewed as a binary polynomial $D_\ell=\sum_{i=0}^{p-2}x_{i, \ell}\cdot x^i\in \mathbb{R}_{p,1}$.
The BR$(p,r)$ requires that $\mathbf{0}^{\mathrm{T}}=H_{BR}\cdot (D_0,D_1,...,D_{p-1})^{\mathrm{T}}$, where $H_{BR}\in \mathbb{R}_{p,1}^{r\times p}$ is the Vandermonde parity-check matrix given by
\begin{equation}\label{eq:001}
	H_{BR}=\left(
	\begin{array}{ccccc}
		1         & 1    & 1     & \cdots & 1        \\
		1   & x & x^2 & \cdots & x^{p-1}   \\
		\vdots    & \vdots & \vdots & \ddots      & \vdots \\
		1 & x^{r-1}& x^{2(r-1)} & \cdots & x^{(r-1)(p-1)} 
	\end{array}
	\right),
\end{equation}
and $\mathbf{0}$ is a zero-row vector.

\begin{figure}[t]
	\centering
	\begin{minipage}{0.49\linewidth}
		\centering
		\includegraphics[width=0.95\textwidth, height=0.5\textwidth]{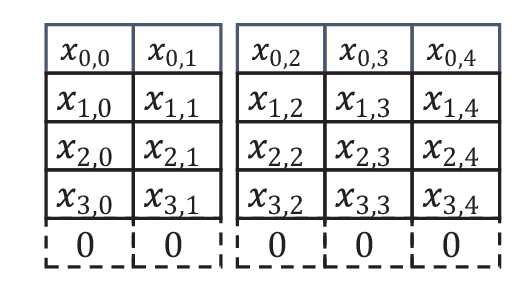}
		\caption{Diagram of the BR code\\with $p=5$ and $r=3$.}
		\label{fig:0}
	\end{minipage}
	\begin{minipage}{0.49\linewidth}
		\centering
		\includegraphics[width=1\textwidth, height=0.5\textwidth]{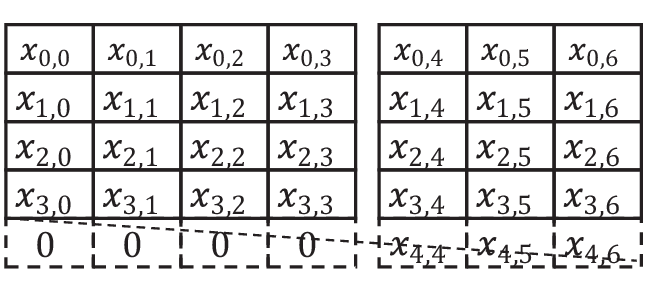}
		\caption{Diagram of the generalized RDP code with $p=5$ and $r=3$.}
		\label{fig:1}
	\end{minipage}
\end{figure}

BR codes have an intuitive graphical representation and Fig.~\ref{fig:0} provides an example of BR$(5, 3)$ to demonstrate it.
In Fig.~\ref{fig:0}, the last row is imaginary to facilitate operations, the leftmost two data columns are information columns of the BR$(5, 3)$, and the rightmost three data columns are all parity columns.
According to the identity $\mathbf{0}^{\mathrm{T}}=H_{BR}\cdot (D_0,D_1,...,D_{p-1})^{\mathrm{T}}$, the result obtained by bit-wise XORing all data columns is an all-zero column.
If each column has been subjected to down-cyclic shifts according to the corresponding column index size, the above result is either an all-zero column or an all-one column. 
This satisfies the need for realization in $\mathbb{R}_{p,1}$, which involves first performing operations in $\mathbb{R}$ and then reducing to $\mathbb{R}_{p,1}$.
The above result is also true if the number of down-cyclic shifts is twice the size of the corresponding column index.
One can know from \cite{blaum1993new} that BR codes are always binary MDS array codes.

\subsection{IP codes}\label{sec:2.b}
IP codes are also constructed in $\mathbb{R}_{p,1}$, but all parity columns are independent of each other, leading to a minimization of the number of parity updates when a data bit is updated~\cite{blaum1995evenodd,blaum1996mds,blaum2019array}.
Precisely, given the prime number $p$ and a positive integer $r$, the IP$(p+r,r)$ is defined as the set of $(p-1)\times (p+r)$ arrays of bits.
In the same way as the BR codes, each column of the array forms a binary polynomial, then the parity-check matrix of the IP$(p+r, r)$ is
$
H_{IP}=\left(
H_{BR}|
I_{r}
\right),
$
where $H_{BR}$ is shown in \eqref{eq:001} and $I_{r}$ is an $r\times r$ identity matrix.
The matrix $H_{IP}$ implies that IP codes also have an intuitive graphical representation similar to that shown in BR codes.
Contrary to BR codes, IP codes are not always binary MDS array codes.
The conditions for making IP codes to be binary MDS array codes can be found in~\cite{blaum1996mds, hou2016mds}.

\subsection{Generalized RDP codes}\label{sec:2.a}

In~\cite{corbett2004row}, the authors presented a binary MDS array code with two parity columns, i.e., RDP codes.
This code was generalized to support more parity columns in \cite{blaum2006family}. 
Generalized RDP codes are not directly constructed by parity-check matrices over $\mathbb{R}_{p,1}$ like the two codes introduced above.
Given a prime number $p$ and a positive integer $r$, the generalized RDP$(p+r-1,r)$ code is defined as the set of $(p-1)\times (p+r-1)$ arrays (denoted by $[x_{i,j}]$, where $x_{i,j}\in \{0,1\}$, the first $p-1$ data columns are information columns, and others are parity columns). 
From~\cite{blaum2006family}, it satisfies the following encoding equations:
\begin{align}
	&x_{i,p-1}=\sum_{j=0}^{p-2}x_{i,j}~\text{for}~0\leq i \leq p-2 \;,\\
	 ~\text{and}~~ &x_{i, p-1+j}=\sum_{\ell=0}^{p-1}x_{i-j\ell, \ell}~\text{for}~{0\leq i\leq p-2 \atop 1 \leq j \leq r-1} \;,
\end{align}
where addition is performed through XOR, all subscripts in the right-hand side of equal signs are modulo $p$, and $x_{p-1,j}=0$ for $j=0,1,...,p-1$.

Similar to BR and IP codes, the generalized RDP codes have an intuitive graphical representation.
Fig.~\ref{fig:1} shows an example of $p=5$ and $r=3$, where the leftmost four data columns are information columns and the last row is imaginary.
Clearly, the first parity column, i.e., $\{x_{i,4}\}_{i=0}^4$, is obtained by bit-wise XORing the first 4 columns. The second parity column, i.e., $\{x_{i,5}\}_{i=0}^4$, is obtained by bit-wise XORing the first 5 columns after each column has been subjected to down-cyclic shifts according to the corresponding column index size.
The third parity column is similar to the second, but the number of down-cyclic shifts in each column becomes twice the corresponding column index size.
The three parity columns of the generalized RDP$(7,3)$ code are obtained by directly deleting the imaginary row.

Generalized RDP codes are not always binary MDS array codes~\cite{blaum2006family}, as are IP codes.
Conditions that make generalized RDP codes to be binary MDS array codes can be found in~\cite{blaum2006family}.
In particular, there is a connection between generalized RDP and IP codes as follows:
\begin{theorem}\label{theorem:01}(\cite{blaum2006family})
	The generalized RDP$(p+r-1, r)$ is a binary MDS array code if the shortened IP$(p+r-1,r)$ with the following parity-check matrix over $\mathbb{R}_{p,1}$ is a binary MDS array code
	\begin{equation}\label{eq:0008}
		H_{SIP}=\left(
		H_{BR}\quad 
		\begin{array}{|cccccc}
			0      & 0      & \cdots & 0      \\
			1      & 0      & \cdots & 0      \\
			0      & 1      & \cdots & 0      \\
			\vdots & \vdots & \ddots & \vdots \\
			0      & 0      & \cdots & 1
		\end{array}
		\right).
	\end{equation}
\end{theorem}

The encoding of the shortened IP code with \eqref{eq:0008} can be analogized from the BR code in Section~\ref{sec:2a}. It is easy to see that the process is similar to that in the generalized RDP$(p+r-1, r)$. 
The only difference is that the latter does not need to calculate the last bit in each parity column and modulo $f_{p,1}(x)$.

\subsection{Notations}\label{sec:2.c}
Throughout this paper, the set $\{0,1,2,3,...\}$ is denoted by $\mathbb{N}$ and the set $\{i,i+1,...,j-1\}$ is denoted by $[i,j)$, where $i\in \mathbb{N},j\in \mathbb{N}$ with $i<j$. 
The transpose of a matrix or vector is marked with the notation $\mathrm{T}$ in the upper right-hand corner.
Unless otherwise stated, suppose that 
\begin{equation}
m=p\tau, \qquad \tau=2^s,
\end{equation}
where $p>1$ is an odd number and $s\in \mathbb{N}$. 
Note that $p$ and $\tau$ are determined if $m$ is given. In addition, $f_{p,\tau}(x)=f_{p,1}^\tau(x)$.


Some special mappings are defined below.
For any $i, j\in \mathbb{N} $ and $a=\sum_{i=0}^{m-1}a_i\cdot x^i\in \mathbb{R}$,  define a mapping
$
\mathcal{A}_{i,j}: \mathbb{R} \rightarrow \mathbb{F}_2^{(m-i)\times (m-j)}
$
by letting $\mathcal{A}_{i,j}(a)$ be the resultant $(m-i)\times (m-j)$ binary matrix after deleting the last $i$ rows and last $j$ columns of the following $m\times m$ binary circulant matrix
\begin{equation}\label{eq:01}
	\left(\begin{array}{ccccc}
		a_{0}   & a_{1} &a_2 & \cdots  & a_{m-1} \\
		a_{m-1} & a_{0} &a_1 & \cdots  & a_{m-2} \\
		\vdots  & \vdots &\vdots & \ddots  & \vdots  \\
		a_{1}   & a_{2} &a_{3} & \cdots & a_{0}
	\end{array}\right).
\end{equation}
That is,
\begin{equation}\label{eq:0+1}
	\mathcal{A}_{i,j}(a)=
	\left(\begin{array}{ccccc}
		a_{0}   & a_{1} &a_2 & \cdots  & a_{m-1-j} \\
		a_{m-1} & a_{0} &a_1 & \cdots  & a_{m-2-j} \\
		\vdots  & \vdots &\vdots & \ddots  & \vdots  \\
		a_{1+i}   & a_{2+i} &a_{3+i} & \cdots & a_{m-j+i}
	\end{array}\right),
\end{equation}
where each subscript is modulo $m$.
From~\cite{subroto2023algebraic}, one can see that $\mathcal{A}_{0,0}$ is an isomorphic mapping.
Moreover, for $i\in \mathbb{N}$, $\mathcal{A}_{i,i}(0)$ is the $(m-i)\times (m-i)$ zero matrix and $\mathcal{A}_{i,i}(1)$ is an $(m-i)\times (m-i)$ identity matrix.
%
Furthermore, for any $\ell_0,\ell_1\in \mathbb{N}$, we define a mapping from the set consisting of all $\ell_0\times \ell_1$ matrices over $\mathbb{R}$ to the set consisting of all $\ell_0 (m-\tau)\times \ell_1(m-\tau)$ matrices over $\mathbb{F}_2$, i.e.,
\begin{equation}\label{eq:7}
	\mathcal{T}_{\ell_0,\ell_1,m}: M_{\ell_0\times \ell_1}(\mathbb{R}) \rightarrow M_{\ell_0 (m-\tau)\times \ell_1(m-\tau)}(\mathbb{F}_2)
\end{equation}
by letting $\mathcal{T}_{\ell_0,\ell_1,m}(B)=\overline{B}$, where $B= [b_{i,j}]\in \mathbb{R}^{\ell_0\times \ell_1}$ and $\overline{B}=[\mathcal{A}_{\tau,\tau}(b_{i,j})] \in \mathbb{F}_2^{\ell_0 (m-\tau)\times \ell_1(m-\tau)}$.

In this paper, the code with $\mathcal{T}_{\ell_0,\ell_1,m}(B)$ as the parity-check matrix has a binary codeword of size $\ell_1\cdot (m-\tau)$, where $\ell_0<\ell_1$.
By default, the codeword is arranged in an $(m-\tau)\times \ell_1$ array of bits in column-first order, and we refer to this code as a binary array code.
In addition, we refer to this code as a binary MDS array code if each codeword array can be restored by any $\ell_1-\ell_0$ columns.

\section{Definitions of Variant codes}\label{sec:3}
This section defines two new classes of binary array codes~(i.e., V-ETBR and V-ESIP codes).
One can see that the generalized RDP codes introduced in Section~\ref{sec:2.a} are a special case of the V-ESIP codes.

To begin with, we define two codes over the polynomial ring $\mathbb{R}_{p,\tau}$ as follows:

\begin{definition}\label{def:03}
	({\bf ETBR Codes}) 
	Let $2\leq r < n$, and {$H=[h_{i,j}]_{0\leq i < r,0\leq j < n}\in \mathbb{R}^{r\times n}$.}
	Define ETBR$(n, r, m=p\tau, H)$ as a code over $\mathbb{R}_{p,\tau}$ determined by the parity-check matrix $H$ that is reduced to over $\mathbb{R}_{p,\tau}$, where each element in $H$ is modulo $f_{p,\tau}(x)$ to be an element over $\mathbb{R}_{p,\tau}$.
\end{definition}

\begin{remark}
	In Definition~\ref{def:03}, the form of $H$ is not fixed and covers $H_{BR}$ in \eqref{eq:001}, so we refer to ETBR$(n, r, m, H)$ as an extended BR code.
\end{remark}

\begin{definition}\label{def:3}
	({\bf ESIP Codes}) 
	Let $n\geq 2, r\geq 2$, and {$H'=[H|\widehat{I}]\in \mathbb{R}^{r\times (n+r-1)}$, where the definition of $H$ is the same as in Definition~\ref{def:03} and $\widehat{I}$ is the matrix after removing the first column of the $r\times r$ identity matrix.}
	Define ESIP$(n,r, m=p\tau, H')$ as a code over $\mathbb{R}_{p,\tau}$ determined by the parity-check matrix $H'$ that is reduced to over $\mathbb{R}_{p,\tau}$, where each element in $H$ is modulo $f_{p,\tau}(x)$ to be an element over $\mathbb{R}_{p,\tau}$.
\end{definition}

\begin{remark}
	In Definition~\ref{def:3}, the ESIP$(n,r, m=p,H')$ is exactly the shortened IP code given by \eqref{eq:0008} if $H=H_{BR}$.
	Obviously, $H'$ has a wider range of parameters so that the ESIP codes can be regarded as an extension of shortened IP codes.
\end{remark}

The variant codes corresponding to ETBR and ESIP codes, i.e., V-ETBR and V-ESIP codes, are defined below. When we refer to ETBR/ESIP codes and V-ETBR/V-ESIP codes as corresponding, it means that they are determined by the same matrix $H$ or $H'$ over $\mathbb{R}$.

%

\begin{definition}\label{def:01}
	({\bf V-ETBR Codes}) 
	Define V-ETBR$(n, r,m=p\tau, H)$ as a binary array code whose parity-check matrix is $\mathcal{T}_{r, n,m}(H)$, where $2\leq r <n$, $\mathcal{T}_{r, n,m}$ is defined in \eqref{eq:7}, and the definition of $H$ is the same as that in Definition~\ref{def:03}. 
\end{definition}
\begin{definition}\label{def:2}
	({\bf V-ESIP Codes}) Define V-ESIP$(n,r, m=p\tau,H')$ as a binary array code whose parity-check matrix is $\mathcal{T}_{r, n+r-1,m}(H')$, where $n\geq 2, r\geq 2$, $\mathcal{T}_{r, n+r-1,m}$ is defined in \eqref{eq:7}, and the definition of $H'$ is the same as that in Definition~\ref{def:3}.
\end{definition}
Conventionally, the last $r$ columns of the array corresponding to the codeword in the above codes are referred to as parity columns and all other columns are referred to as information columns.
We have the following relationship.

{
	\begin{lemma}\label{lemma:01}
		Let $H'$ in Definition~\ref{def:2} be determined by a Vandermonde matrix $H$ such that $h_{1,j}=x^{p-j}$ and $h_{i,j}=h_{1,j}^i$ for $2\leq i < r, 0\leq j <p$, and $p$ is a prime number.
		Then V-ESIP$(p, r, m=p, H')$ is exactly the generalized RDP$(p+r-1, r)$ described in Sec.~\ref{sec:2.a}.
		\begin{proof}
			From Definition~\ref{def:2}, $\mathcal{T}_{r,p+r-1,p}(H')$ is the parity-check matrix of the V-ESIP$(p, r, m=p, H')$ and is given by \eqref{eq:15} at the top of this page, where all unspecified entries are zero.

			\begin{figure*}[t]
			\begin{equation}\label{eq:15}
				\mathcal{T}_{r,p+r-1,p}(H')=\left(
				\begin{array}{cccccccc}
					I_{p-1}&	I_{p-1}&	I_{p-1}&	\cdots&	I_{p-1}&	& &\\
					I_{p-1}&\mathcal{A}_{1,1}(x^{p-1})&	\mathcal{A}_{1,1}(x^{p-2})&	\cdots&	\mathcal{A}_{1,1}(x)&	I_{p-1}&&	\\
					\vdots&\vdots&\vdots&	\ddots&	\vdots&	&\ddots&\\
					I_{p-1}&\mathcal{A}_{1,1}(x^{(p-1)(r-1)})&	\mathcal{A}_{1,1}(x^{(p-2)(r-1)})&\cdots&	\mathcal{A}_{1,1}(x^{r-1})&	&	&I_{p-1}\\
				\end{array}
				\right).
			\end{equation}
			\hrulefill
			\end{figure*}

			Let $\mathbf{b}_0,\mathbf{b}_1,...,\mathbf{b}_{p-2}\in \mathbb{F}_2^{p-1}$ denote all $p-1$ information columns in the codewrod.
			We next show that any parity column generated by the V-ESIP code is the same as that in the generalized RDP code described in Sec.~\ref{sec:2.a}.
			
			Let $\mathbf{b}_{p-1},\mathbf{b}_{p},...,\mathbf{b}_{p+r-2}\in \mathbb{F}_2^{p-1}$ denote all $r$ parity columns of the V-ESIP code.
			One can easily know from \eqref{eq:15} that $\mathbf{b}_{p-1}$ is obtained by bit-wise XORing of all $p-1$ information columns.
			For any $i\in [1,r)$, the $i$-th parity column of the V-ESIP code is obtained by
			\begin{equation}\label{eq:101}
				\mathbf{b}_{p-1+i}^{\mathrm{T}}=\sum_{j = 0}^{p-1}\mathcal{A}_{1,1}(x^{(p-j)i}) \cdot \mathbf{b}_j^{\mathrm{T}}=\sum_{j = 0}^{p-1}\mathcal{A}_{1,0}(x^{(p-j)i}) \cdot (\mathbf{b}_j,0)^{\mathrm{T}}.
			\end{equation} 
			Note that calculating $\mathcal{A}_{1,0}(x^{(p-j)i})\cdot (\mathbf{b}_j, 0)^{\mathrm{T}}$ is equivalent to removing the last element from the result of $\mathcal{A}_{0,0}(x^{(p-j)i})\cdot (\mathbf{b}, 0)^{\mathrm{T}}$.
			Furthermore, $\mathcal{A}_{0,0}(x^{(p-j)i})=\left( \mathcal{A}_{0,0}(x^{p-j})\right)^i$, where $\mathcal{A}_{0,0}(x^{p-j})$ can be regarded as the operator of performing $j$ times down-cyclic shift on a vector.
			Assume that each data column has an imaginary bit attached at the end,
			thus, \eqref{eq:101} indicates that each $\mathbf{b}_{p-1+i}, i\in [1,r)$ can be obtained by bit-wise XORing of the first $p$ columns after each column has been subjected to down-cyclic shifts according to $i$ times the corresponding column index size.
			Each parity column needs to remove the last bit in the result.
			The above process is consistent with the graphical representation of the generalized RDP code, as shown in Fig.~\ref{fig:1}.
			This completes the proof.		
		\end{proof}
	\end{lemma}
	Lemma~\ref{lemma:01} explicitly provides a binary parity-check matrix for the generalized RDP$(p+r-1, r)$.
	From the perspective of the binary parity-check matrix, all fast computations about the generalized RDP codes, such as those proposed in \cite{huang2016improved,hou2018unified}, can thus be regarded as scheduling schemes for matrix operations over binary fields.
	Furthermore, any existing scheduling algorithm for general matrix operations over binary fields may be used to accelerate the computation of generalized RDP codes, such as~\cite{plank2012heuristics,huang2007optimizing}.
	
	Recall that Theorem~\ref{theorem:01} established a connection between generalized RDP codes and shortened IP codes, which can be viewed as a special case of the connection between V-ESIP codes and ESIP codes.
	It remains an open problem whether there exists a general connection between V-ESIP and ESIP codes, as well as between V-ETBR and ETBR codes.
	The next section is devoted to these issues.
}

\section{Conditions that make variant codes binary MDS array codes}\label{sec:iv}
This section proposes the conditions that make the variant codes ({see Definitions~\ref{def:01} and \ref{def:2}}) binary MDS array codes by exploring the general connections between them and their counterparts over polynomial rings ({see Definitions~\ref{def:03} and \ref{def:3}}).
TABLE~\ref{tab:1} defines some important symbols to be used later.

\subsection{Rank of the square matrix $\mathcal{T}_{\ell, \ell,m}(V)$}
We first explore the the rank of $\mathcal{T}_{\ell, \ell,m}(V)$.
The following lemmas are useful.

\begin{lemma}\label{lemma:3}
	{
		Assume that $\ell_0\geq 2,\ell_1\geq 2$, $x^\tau+1|a_{i,j}+a_{i,k}$ with $a_{i,j},a_{i,k}\in \mathbb{R}, i\in [0,\ell_0), j,k\in [0, \ell_1)$.
		Let each $\mathbf{a}_{i,j}$ denote the binary coefficient vector of $a_{i,j}$, i.e., $a_{i,j}=\mathbf{a}_{i,j}\cdot (1,x,...,x^{m-1})^{\mathrm{T}}$, and let $\overline{\mathbf{a}}_{i,j}$ denote the vector after $\mathbf{a}_{i,j}$ deletes the last $\tau$ elements.
		If all vectors in the set $\{\left(\overline{\mathbf{a}}_{i,0}, ...,\overline{\mathbf{a}}_{i,\ell_1-1} \right)\in \mathbb{F}_{2}^{1\times (m-\tau)\ell_1}\}_{i=0}^{\ell_0-1}$ are $\mathbb{F}_2$-linearly dependent, i.e., $\sum_{i=0}^{\ell_0-1}c_i\cdot \left(\overline{\mathbf{a}}_{i,0}, ...,\overline{\mathbf{a}}_{i,\ell_1-1} \right)=\mathbf{0}_{1\times (m-\tau)\ell_1}$, where each $c_i\in \mathbb{F}_2$ and $c_0,c_1,...,c_{\ell_0-1}$ are not all zero,
		then the result of $\sum_{i=0}^{\ell_0-1}c_i\cdot \left({\mathbf{a}}_{i,0}, ...,{\mathbf{a}}_{i,\ell_1-1} \right)$ must have the form of 
		$
		(\mathbf{0}_{1\times (m-\tau)},\mathbf{u} | \mathbf{0}_{1\times (m-\tau)}, \mathbf{u}, | ... | \mathbf{0}_{1\times (m-\tau)},\mathbf{u}),
		$
		where $\mathbf{u}\in \mathbb{F}_2^{1\times \tau}$.
	}	
	\begin{proof}
		From the condition, we immediately have
		\begin{equation} 
			\begin{aligned}
			&\sum_{i=0}^{\ell_0-1}c_i\cdot \left({\mathbf{a}}_{i,0}, ...,{\mathbf{a}}_{i,\ell_1-1} \right)\\
		=&	(\mathbf{0}_{1\times (m-\tau)},\mathbf{u}_0 | \mathbf{0}_{1\times (m-\tau)}, \mathbf{u}_1, | ... | \mathbf{0}_{1\times (m-\tau)},\mathbf{u}_{\ell_1-1})
			\end{aligned}
		\end{equation} 
		where each $\mathbf{u}_i\in \mathbb{F}_2^{1\times \tau}$.
		In the above formula, the sum of any two parts is $(\mathbf{0}_{1\times (m-\tau)}, \mathbf{u}_j+\mathbf{u}_k)=\sum_{i=0}^{\ell_0-1}c_i\cdot (\mathbf{a}_{i,j}+\mathbf{a}_{i,k})$, where $0\leq j<k<\ell_1$.
		Since $x^\tau + 1 | a_{i,j}+a_{i,k}$, then the sum $(\mathbf{0}_{1\times (m-\tau)}, \mathbf{u}_j+\mathbf{u}_k)$ is a binary coefficient vector of the polynomial that is a multiple of $x^\tau+1$.
		However, $\mathbf{u}_j+\mathbf{u}_k$ contains only $\tau$ elements. This results in $\mathbf{u}_j+\mathbf{u}_k$ having to be a zero vector.
		Therefore, we have $\mathbf{u}_j= \mathbf{u}_k$ with $j\neq k$.
		This completes the proof.		
	\end{proof}
	
\end{lemma}

{
	\begin{remark}\label{re:3}
		From the proof of Lemma~\ref{lemma:3}, one can readily know that $\mathbf{u}=\mathbf{0}_{1\times \tau}$ in Lemma~\ref{lemma:3}, if $x^\tau+1| a_{i,j},  i\in [0,\ell_0), j \in [0,\ell_1)$.
	\end{remark}
}

%
\begin{table*}[]
	\centering
	\caption{Important symbols used in Section~\ref{sec:iv}\label{tab:1}}
	\begin{tabular}{c||l}
		\hline
		Symbol                         & \multicolumn{1}{c}{Definition}  \\ \hline \hline
		$\ell$                         & a positive number not less than two.                                               \\          
		$V$                            & $V=[v_{i,j}]$  is an $\ell \times \ell $ square matrix over $\mathbb{R}$.                                                 \\
		$\mathcal{B}(i, j)$                    & $ \mathcal{B}(i, j)=\left(\mathcal{A}_{\tau,0}(v_{i,0}),    ..., \mathcal{A}_{\tau,0}(v_{i,j-1})\right)$,  $i\in [0, \ell), j \in [1, \ell+1)$.                                                             \\
		$\mathcal{B}_\tau(i, j)$           & $\mathcal{B}_\tau(i, j)=\left(\mathcal{A}_{\tau,\tau}(v_{i,0}),  ..., \mathcal{A}_{\tau,\tau}(v_{i,j-1})\right)$, $i\in [0, \ell), j \in [1, \ell+1)$.   \\
		$\overline{\mathbf{v}}_{i, j}$ & a non-zero codeword with vector form generated by generator matrix $\mathcal{B}_\tau(i, j)$.                        \\
		${\mathbf{v}}_{i, j}$          & the non-zero codeword with vector form generated by generator matrix $\mathcal{B}(i, j)$ and corresponds to $\overline{\mathbf{v}}_{i, j}$.\\ \hline
	\end{tabular}
\end{table*}


\begin{lemma}\label{lemma:4.1}
	
	The square matrix $\mathcal{T}_{\ell, \ell,m }(V)$ has full rank if the following conditions are satisfied:
	{
		\begin{enumerate}[1)]
			\item $V$ has full rank over $\mathbb{R}_{p,\tau}$, i.e., $\gcd(|V|, f_{p,\tau}(x))=1$, where $|V|$ is the determinant of $V$.
			\item For any $ 0\leq i < \ell$, then $\mathcal{B}_\tau(i,\ell)$ in TABLE~\ref{tab:1} has full row rank over $\mathbb{F}_2$.
			\item\label{lm:3)} For any $0\leq i < \ell, 0\leq j < \ell$, then $x^\tau + 1 | v_{i,j}$.
		\end{enumerate}
		When $v_{0,j}=1, \forall j\in[0,\ell)$, \ref{lm:3)}) is relaxed to 
		\begin{enumerate}[3')]
			\item For any $1\leq i < \ell, 0\leq j<k < \ell$, then $x^\tau + 1 | v_{i,j}+v_{i,k}$.
		\end{enumerate}
	}
	
	\begin{proof}
		According to TABLE~\ref{tab:1}, $\mathcal{T}_{\ell, \ell,m }(V)$ is composed of $\mathcal{B}_{\tau}(0,\ell),\mathcal{B}_{\tau}(1,\ell),...,\mathcal{B}_{\tau}(\ell-1,\ell)$.
		Since each $\mathcal{B}_{\tau}(i,\ell), i\in [0, \ell)$, has full row rank, we only need to prove that there is no $\overline{\mathbf{v}}_{0,\ell}, \overline{\mathbf{v}}_{ 1,\ell},   ..., \overline{\mathbf{v}}_{\ell-1,\ell}$, which are $\mathbb{F}_2$-linearly dependent.
		By contradiction, assume that there exists $\overline{\mathbf{v}}_{0,\ell}, \overline{\mathbf{v}}_{ 1,\ell},  ..., \overline{\mathbf{v}}_{ \ell-1,\ell}$ such that they are $\mathbb{F}_2$-linearly dependent, i.e., $\sum_{i=0}^{\ell-1}c_i\overline{\mathbf{v}}_{i,\ell}= \mathbf{0}_{1\times (m-\tau)\ell}$ where each $c_i \in \mathbb{F}_2$ and $c_0,c_1,...,c_{\ell-1}$ are not all zero.
		
		We first consider the third condition of $x^\tau+1|v_{i,j}$.
		According to Remark~\ref{re:3} and the facts that $x^\tau+1| v_{i,j} $, each ${\mathbf{v}}_{ i,\ell}$ is the vector consisting of the binary coefficient vectors in $q_{i,\ell}\cdot (v_{i,0}, v_{i,1}, ..., v_{i,\ell-1})$, where $q_{i,\ell}\in \mathbb{R}\setminus \{0\}$ and $\deg(q_{i,\ell})<m-\tau$,	
		then
		$\sum_{i=0}^{\ell-1}c_i {\mathbf{v}}_{ i,\ell}=	(\mathbf{0}_{1\times m},...,\mathbf{0}_{1\times m})$.
		Therefore, we have $\sum_{i=0}^{\ell-1}c_iq_{i,\ell}\cdot  v_{i,j} =0 \mod x^m+1$ for $j\in [0, \ell)$.
		By taking $j=0,1,...,\ell-1$, the above equations can be converted into
		\begin{equation}\label{eq:13}
			\Gamma_0\cdot 
			\begin{pmatrix}
				c_0\cdot q_{0,\ell}\\
				c_1\cdot q_{1,\ell}\\
				\vdots\\
				c_{\ell-1}\cdot q_{\ell-1,\ell}
			\end{pmatrix}=\mathbf{0}^{\mathrm{T}}, 	
		\end{equation}
		where $\mathbf{0}$ is a zero-row vector and
		\begin{equation}
		\Gamma_0=		\left(
		\begin{array}{cccc}
			v_{0,0}     & v_{1,0}       & \cdots & v_{\ell-1,0}    \\
			v_{0,1}      & v_{1,1}       & \cdots & v_{\ell-1,1}   \\
			\vdots & \vdots       & \ddots & \vdots            \\
			v_{0,\ell-1}      &  v_{1,\ell-1}        & \cdots & v_{\ell-1,\ell-1}  
		\end{array}\right).
		\end{equation}
	 In \eqref{eq:13}, all operations are performed in $\mathbb{R}$.
		Note that each $c_i\in \mathbb{F}_2$ and $\deg(q_{i,\ell})<m-\tau$,
		we can solve the above linear equations in $\mathbb{R}_{p,\tau}$. 
		Since $|\Gamma_0| = |V|$ is invertible over $\mathbb{R}_{p,\tau}$, then $c_0q_{0,\ell},...,c_{\ell-1}q_{\ell-1,\ell}$ in \eqref{eq:13} must all be zero according to Cramer's rule.
		Moreover, each $q_{i,\ell}\neq 0$ with $\deg(q_{i,\ell})<m-\tau$, so that $c_0=c_1=\cdots=c_{\ell-1}=0$.
		This contradicts the assumption at the beginning.
		
		Consider the third condition of $x^\tau + 1 | v_{i,j}+v_{i,k}$ instead, then Lemma~\ref{lemma:4.1} gives
		$\sum_{i=0}^{\ell-1}c_i {\mathbf{v}}_{ i,\ell}=(\mathbf{0}_{1\times (m-\tau)},\mathbf{u} |  ... | \mathbf{0}_{1\times (m-\tau)},\mathbf{u})$, where $\mathbf{u}\in \mathbb{F}_2^{m-\tau}$.
		Since $v_{0,j}=1, \forall j\in [0,\ell)$, we have 
		\begin{equation}\label{eq:14211}
			\begin{aligned}
		&\sum_{i=1}^{\ell-1}c_i {\mathbf{v}}_{ i,\ell}\\
		=&c_0\cdot \mathbf{v}_{0,\ell} + (\mathbf{0}_{1\times (m-\tau)},\mathbf{u} |  ... | \mathbf{0}_{1\times (m-\tau)},\mathbf{u})= (\mathbf{u}' |  ... |\mathbf{u}'),
			\end{aligned}
		\end{equation}
		where $\mathbf{u}'\in \mathbb{F}_2^m$.
		Based on the fact that any two part of the form in \eqref{eq:14211} sum to zero, we have $\sum_{i=1}^{\ell-1}c_iq_{i,\ell}\cdot ( v_{i,j}+v_{i,k} )=0 \mod x^m+1$ for $0\leq j<k<\ell $.
		By taking $(j,k)=(0,1),(0,2),...,(0,\ell-1)$, the above equations can be converted into
		\begin{equation}\label{eq:5.31}
			\Gamma_1 \cdot
			\begin{pmatrix}
				c_1\cdot q_{1,\ell}\\
				c_2\cdot q_{2,\ell}\\
				\vdots\\
				c_{\ell-1}\cdot q_{\ell-1,\ell}
			\end{pmatrix} =\mathbf{0}^{\mathrm{T}}, 
		\end{equation}
		where $\mathbf{0}$ is a zero-row vector and
	\begin{equation} 
	\Gamma_1=
		\begin{pmatrix}
			v_{1,0} + v_{1,1}      & \cdots & v_{\ell-1,0} + v_{\ell-1,1}      \\
			v_{1,0}+ v_{1,2}       & \cdots & v_{\ell-1,0} + v_{\ell-1,2}      \\
			\vdots                 & \ddots & \vdots                           \\
			v_{1,0} + v_{1,\ell-1} & \cdots & v_{\ell-1,0} + v_{\ell-1,\ell-1} \\
		\end{pmatrix}.
	\end{equation}
		In \eqref{eq:5.31}, all operations are performed in $\mathbb{R}$.
		Note that each $c_i\in \mathbb{F}_2$ and $\deg(q_{i,\ell})<m-\tau$,
		we can solve the above linear equations in $\mathbb{R}_{p,\tau}$. 
		One can know that the determinant of $\Gamma_1$ is equal
		\begin{equation}
			\begin{aligned}
				& \left|
				\begin{array}{cccc}
					1      & 0                      & \cdots & 0                                \\
					1      & v_{1,0}+ v_{1,1}       & \cdots & v_{\ell-1,0} + v_{\ell-1,1}      \\
					\vdots & \vdots                 & \ddots & \vdots                           \\
					1      & v_{1,0} + v_{1,\ell-1} & \cdots & v_{\ell-1,0} + v_{\ell-1,\ell-1} \\
				\end{array}
				\right|   \\ 
				= & \left|
				\begin{array}{cccc}
					1      & v_{1,0}      & \cdots & v_{\ell-1,0}      \\
					1      & v_{1,1}      & \cdots & v_{\ell-1,1}      \\
					\vdots & \vdots       & \ddots & \vdots            \\
					1      & v_{1,\ell-1} & \cdots & v_{\ell-1,\ell-1}
				\end{array}
				\right|=|V|,
			\end{aligned}
		\end{equation}
		where the first equality is obtained by subtracting the appropriate multiple of the first column from all other columns.
		Thus,
		$|\Gamma_1| = |V|$ is also invertible over $\mathbb{R}_{p,\tau}$. Then $c_1q_{1,\ell},...,c_{\ell-1}q_{\ell-1,\ell}$ in \eqref{eq:13} must all be zero according to Cramer's rule.
		Similarly, note that each $q_{i,\ell}\neq 0$ with $\deg(q_{i,\ell})<m-\tau$, so we must have that $c_1=\cdots=c_{\ell-1}=0$, then $c_0=0$.
		This contradicts the assumption at the beginning.
		This completes the proof.
	\end{proof}
	
\end{lemma}

The above lemma reveals the connection between the ranks of $V$ and $\mathcal{T}_{\ell, \ell,m}(V)$.
In Lemma~\ref{lemma:4.1}, the latter two conditions are easily met.
More precisely, the third condition only requires that any $v_{i,j}$ is a multiple of $x^\tau+1$, and the second condition can be satisfied by the following lemma, which is easily obtained through the proof of Proposition 6 in \cite{lv2023new}.
\begin{lemma}\label{le:4}(\cite{lv2023new})
Let $a,b\in \mathbb{R}$, then $\mathcal{A}_{\tau,\tau}(a)$ has full row rank over $\mathbb{F}_2$ if $\gcd(a, x^m+1)=x^\tau+1$; $\mathcal{A}_{\tau,\tau}(a, b)$ has full row rank over $\mathbb{F}_2$ if $\gcd(a+b, x^m+1)=x^\tau+1$.
\end{lemma}


\subsection{Conditions for binary MDS array codes}\label{sec:4.b}
We now present the conditions that make the variant codes binary MDS array codes, by establishing the connections between them and the corresponding codes over polynomial rings.
To begin with, the following theorem on V-ETBR codes can be obtained.
\begin{theorem}\label{theorem:02}
	V-ETBR$(n, r, m=p\tau, H)$ is a binary MDS array code if 
	\begin{enumerate}[1)]
		\item The corresponding ETBR$(n, r,m, H)$ is an MDS code over $\mathbb{R}_{p,\tau}$.		
		\item {For any $0\leq i < r, 0\leq j < n$, then $\gcd(h_{i,j}, x^m+1)=x^\tau+1$.}
	\end{enumerate}
	When $h_{0,j}=1, \forall j\in[0,n)$, the above last condition is replaced with
	\begin{enumerate}[2')]
		\item For any $1\leq i < r, 0\leq j,k< n$ and $j\neq k$, then $\gcd(h_{i,j}, x^m+1)=x^\tau+1$ or $\gcd(h_{i,j}+h_{i,k}, x^m+1)=x^\tau+1$.
	\end{enumerate}

	
	\begin{proof}
		We only need to prove that any $\mathcal{T}_{\ell \times \ell}(V)$ for $\ell= r$ has full rank, where all elements in $V$ are determined by $H$.
		This is easily derived from Lemmas~\ref{lemma:4.1} and \ref{le:4}.		
		
	\end{proof}
\end{theorem}

The following theorems on V-ESIP codes can be obtained.
\begin{theorem}\label{theorem:2}
	When the first row of $H$ in $H'$ is an all-one row, the V-ESIP$(n, r,m=p\tau, H')$ is a binary MDS array code if 
	\begin{enumerate}[1)]
		\item The corresponding ESIP$(n,r,m, H')$ is an MDS code over $\mathbb{R}_{p,\tau}$.
		\item For any $1\leq  i < r$ and $0\leq j<k<n$, then $x^\tau+1|h_{i,j}+h_{i,k}$.
	\end{enumerate}
	\begin{proof}
		Without loss of generality, we only need to prove $\mathcal{T}_{\ell, \ell,m}(V)$ for any $1<\ell\leq r$ has full rank, where elements in $\mathcal{T}_{\ell, \ell,m}(V)$ are determined by $H'$ and $\{v_{0,j}=1\}_{j=0}^{\ell-1}$. We prove this via Lemma~\ref{lemma:4.1}.
		First, the first condition of Lemma~\ref{lemma:4.1} is satisfied since the corresponding ESIP$(n,r,m, H')$ is an MDS code over $\mathbb{R}_{p,\tau}$.
		Furthermore, the fact that $V$ with $\ell=2$ have full rank over $\mathbb{R}_{p,\tau}$ leads to $\gcd(h_{i,j}+h_{i,k}, f_{p,\tau}(x))=1, 1\leq  i < r, 0\leq j<k<n$.
		Recall that the condition of $x^\tau+1|h_{i,j}+h_{i,k}$, then we have $\gcd(h_{i,j}+h_{i,k}, x^m+1)=x^\tau+1$.
		One can easily see from Lemma~\ref{le:4} that the second condition of Lemma~\ref{lemma:4.1} is thus satisfied.
		The latter third condition of Lemma~\ref{lemma:4.1} is obviously satisfied. 
		This completes the proof.
	\end{proof}
\end{theorem}

\begin{remark}
	Now, the correctness of Theorem~\ref{theorem:01} can be readily proven by Theorem~\ref{theorem:2}, just by setting $\tau = 1$.
	Theorem~\ref{theorem:01} requires $p$ to be an odd prime number for shortened IP codes.
	Theorem~\ref{theorem:2} provides additional clarification by demonstrating that $p$ only needs to be odd.
\end{remark}

{
	In Theorem~\ref{theorem:2}, the rightmost end of $H'$ is not necessarily an identity matrix.
	Since the existence of an identity matrix can simplify encoding, we consider the following case that does not require the first row of $H$ to be an all-one row (only the last column to be constrained).
}

\begin{theorem}\label{theorem:3}
	{
		When the rightmost end of $H'$ is an $r\times r$ identity matrix, i.e.,  the last column of $H$ is $(1,0,0,...,0)^{\mathrm{T}}$,
		then the V-ESIP$(n, r,m=p\tau, H')$ is a binary MDS array code if 
		\begin{enumerate}[1)]
			\item The corresponding ESIP$(n,r,m, H')$ is an MDS code over $\mathbb{R}_{p,\tau}$.
			\item For any $0\leq  i < r$ and $0\leq j<n-1$, then $\gcd(h_{i,j}, x^m+1)=x^\tau+1$.
		\end{enumerate}
	}
	\begin{proof}
		Without loss of generality, we only need to prove $\mathcal{T}_{\ell, \ell,m}(V)$ for any $1\leq \ell\leq r$ has full rank, where elements in $\mathcal{T}_{\ell, \ell,m}(V)$ are determined by 
		{$H$ after removing the last column.
		}
		Similarly, we prove this via Lemma~\ref{lemma:4.1}.
		First, the first condition of Lemma~\ref{lemma:4.1} is satisfied since the corresponding ESIP$(n,r,m, H')$ is an MDS code over $\mathbb{R}_{p,\tau}$.
		Lemma~\ref{le:4} and $\gcd(h_{i,j}, x^m+1)=x^\tau+1$ lead to that the second condition of Lemma~\ref{lemma:4.1} holds.
		Finally, the {former} third condition of Lemma~\ref{lemma:4.1} obviously holds. 
		This completes the proof.
	\end{proof}
\end{theorem}

\section{Explicit constructions \& fast computations }\label{sec:3.b}
Based on the conditions given in Sec.~\ref{sec:4.b}, we next present some explicit constructions for the V-ETBR/V-ESIP binary MDS array codes.
In particular, Vandermonde matrices and Cauchy matrices are two classes of matrices commonly used in the construction of MDS codes.
They both have a regular structure, and their determinants can be easily calculated.
By setting appropriate entries, one can easily make sub-matrices of the Cauchy-based/Vandermonde-based parity-check matrix having full rank.
For more details, please refer to \cite{schindelhauer2013maximum,blomer1995xor,yu2023reed}.
This paper also explores the use of the two matrices in our constructions.

\subsection{Constructions}\label{sec:V.1}

To begin with, suppose that $f_{p,1}(x)$ in \eqref{eq:02} can be completely factorized into 
$
f_{p,1}(x)=f_0(x)\cdot f_1(x)\cdots f_{\mu - 1}(x),
$
where each $f_i(x)$ is an irreducible polynomial over $\mathbb{F}_2[x]$ and $\lambda = deg(f_0(x)) \leq deg(f_1(x)) \leq \cdots \leq deg(f_{\mu - 1}(x))$. 
Note that $\lambda = p - 1$ if $2$ is a primitive element in $p$-ary finite field $\mathbb{F}_p$~\cite{blaum1996mds}.
Then, we have the following construction for the V-ESIP MDS array codes with any number of parity columns, based on Cauchy matrices.

\begin{construction}\label{cons:1}
	({\bf V-ESIP MDS} array codes with $r\geq 2$) 
	{
		Let $\{a_0,...,a_{r-1}\}$ and $\{b_0,...,b_{n-2}\}$ are two sets of elements from $\mathbb{R}$, where $\deg(a_i)<\lambda, \deg(b_j)<\lambda$ and $a_i\neq b_j$ for any $i,j$, then
		the
		V-ESIP$(n, r\geq 2,m=p\tau, H'=[H_I|I_{r\times r}])$ is a binary MDS array code, where $H_I=[({x^\tau+1})\cdot g_{i,j}]\in \mathbb{R}^{r\times (n-1)}$ and $g_{i,j}$ denotes the inverse of ${a_i+b_j}$ over $\mathbb{R}_{p,\tau}$ {that always exists due to the degree of ${a_i+b_j}$ less than $\lambda$.}
	}
	\begin{proof}
		We prove this via Theorem~\ref{theorem:3}.
		Let $H_I'=[g_{i,j}=\frac{1}{a_i+b_j}]\in \mathbb{R}_{p,\tau}^{r\times (n-1)}$ that is a Cauchy matrix. 
		Obviously, the determinant of any square sub-matrix of $H_I'$ is invertible over $\mathbb{R}_{p,\tau}$, since it is the product of some elements in the sets $\{a_i+a_j\}_{i\neq j}, \{b_i+b_j\}_{i\neq j}, \{\frac{1}{a_i+b_j}\}$~\cite{blomer1995xor}, where any $a_i, b_j$ has the degree less than $\lambda$.
		Note that the determinant of the corresponding square sub-matrix of $H_I$ is $(x^\tau+1)^\xi$ times the above result for some $\xi\geq 1$, and $\gcd(x^\tau+1, f_{p,\tau}(x))=1$.
		This results in the determinant of any square sub-matrix of $H_I$ having to be invertible over $\mathbb{R}_{p,\tau}$.
		Thus, the first condition of Theorem~\ref{theorem:3} is satisfied.
		Furthermore, we have $\gcd({(x^\tau+1)}g_{i,j}, f_{p,\tau}(x))=1$, leading to the second condition of Theorem~\ref{theorem:3} holds.
		This completes the proof.
	\end{proof}
\end{construction}

%

\begin{remark}
%
	To our knowledge, many works on MDS codes seek efficient computation by mapping Cauchy-based parity-check matrices over finite fields to binary matrices~\cite{blomer1995xor,plank2006optimizing,plank2007jerasure,tang2021fast}. This enables the use of scheduling algorithms of binary matrix-vector multiplication, reducing the number of operations.
Construction 1 changes finite fields to polynomial rings and also provides a binary mapping.
	Existing scheduling algorithms of binary matrix-vector multiplication may be applicable to the variant codes in Construction 1, such as those proposed in \cite{plank2006optimizing,plank2007jerasure,tang2021fast}. 
	Notably, our mapping is more convenient than the previous one to design and analyze the number of 1s in the resulting matrix after mapping, as it is obtained through circulant matrices, while the other is by taking the modulus of an irreducible polynomial.
	The new mapping offers a new idea for developing efficient scheduling algorithms for Cauchy-based codes.
\end{remark}

Based on Vandermonde matrices, Construction~\ref{cons:2} provides the construction of the V-ETBR MDS array codes with any number of parity columns.
This construction can also be found in \cite{lv2023new}, but a different proof is provided. 
Since the proof is based on Theorem~\ref{theorem:02}, which reveals the general connection between variant codes and codes over polynomial rings (not specified in \cite{lv2023new}), we only need to check whether the associated matrices over polynomial rings satisfy two simple conditions.
This results in a proof process that is more concise and efficient compared to the one in \cite{lv2023new} (which focuses directly on binary parity-check matrices).
In addition, this proof shows the wide applicability of Theorem~\ref{theorem:02}.
From the proof of any construction proposed in this paper, one can easily see that the codes over $\mathbb{R}_{p,\tau}$ corresponding to the variant codes are also MDS codes.

To simplify the representation of elements in the Vandermonde matrix $H$, we let $h_{0,i}=1,\forall i\in [0,n),$ and
$
h_{i}:=h_{1,i}, \forall i\in [0, n),
$
such that $h_{j,i}=h_i^j, \forall j\in [1, r), i \in [0,n)$.

\begin{construction}\label{cons:2}
	({\bf V-ETBR MDS} array codes with $r\geq 2$) 
	Let $H\in \mathbb{R}^{r\times n}$ be a Vandermonde matrix, $n=2^{n_0}, n_0\leq \lambda,$ and $h_i= (1+x^\tau)\cdot h_i', \forall i\in [0, n)$, where $\{h_i'\}_{0\leq i < n}$ is given by
	$
	h_0'=0$ and $h_{i+2^j}' = h_i'+x^j,  {0\leq j <n_0,  0\leq i < 2^j}.
	$
	Then, the V-ETBR$(n, 2\leq  r < n,m=p\tau, H)$ is a binary MDS array code.
	
	\begin{proof}
		We prove this via Theorem~\ref{theorem:02}. 
		Since the degree of any $h_i'$ is less than $\lambda$,
		we have $gcd(h_i^j, x^m+1 )=(x^\tau+1) \cdot gcd( (h_i')^j,  f_{p,1}^\tau(x))=x^\tau+1$, where $1\leq i < r$ and $0\leq j < k<n$. This results in that the latter second condition of Theorem~\ref{theorem:02} holds.
		For the first condition of Theorem~\ref{theorem:02}, we have $gcd(h_j+h_k, f_{p,\tau}(x))= gcd(h_j'+h_k', f_{p,1}^\tau(x))=1$, where $0\leq j<k<n$. Then any $r\times r$ Vandermonde sub-matrix of $H$ is invertible over $\mathbb{R}_{p,\tau}$, leading to the ETBR$(n, r, m=p\tau, H)$ being MDS code over $\mathbb{R}_{p,\tau}$.
		This completes the proof.
	\end{proof}
\end{construction}

\begin{remark}
	In \cite{lv2023new}, the authors provided a fast scheduling scheme for the syndrome computation of Construction~\ref{cons:2} with $2\leq r\leq 3$.
	The next subsection (i.e., Section~\ref{sec:v.a}) will propose its generalization to be suitable for any $r\geq 2$.
\end{remark}

According to Theorem~\ref{theorem:2}, it is not difficult to check that the V-ESIP$(n, r=3,m=p\tau, H')$ is a binary MDS array code if $H'$ has the same $H$ as Construction~\ref{cons:2}.
The following provides the Vandermonde-based construction for the V-ESIP MDS array code with $r=4$.

\begin{construction}\label{cons:3}
	({\bf V-ESIP MDS} array codes with $r=4$)  
	Let $H\in \mathbb{R}^{r\times n}$ be a Vandermonde matrix, $n=2^{n_1} + 1, n_1\leq w = \lfloor \frac{\lambda-1}{2} \rfloor$, $h_{n-1}=0$, and $h_i=(h_i'+x^{w})\cdot (1+x^\tau)$,
	where $i\in [0, 2^{n_1})$ and $\{h_i'\}_{0\leq i < 2^{n_1}}$ is given by	
	$
	h_0'=0,  h_{i+2^j}' = h_i'+x^j,  {0\leq j < n_1,  0\leq i < 2^j}.
	$
	Then, the V-ESIP$(n,r=4,m=p\tau, H')$ is a systematic binary MDS array code.

	\begin{proof}
		We prove this via Theorem~\ref{theorem:2}.
		The second condition in Theorem~\ref{theorem:2} obviously holds. 
		For the first condition in Theorem~\ref{theorem:2}, we only need to prove that any $4\times 4$ sub-matrix of $H'$ is invertible over $\mathbb{R}_{p,\tau}$.
		Specifically, we first consider any $4\times 4$ sub-matrix of $H$ in $H'$, which is a Vandermonde square matrix.
		Clearly, for any $0\leq i<j<n-1$, we have that {$gcd(h_i+h_j, f_{p,\tau}(x))=gcd(h_i'+h_j', f_{p,1}^\tau(x))$} and $gcd(h_i+h_{n-1}, f_{p,1}^{\tau}(x))=gcd(h_i'+x^{w}, f_{p,1}^{\tau}(x))$.
		Since each $deg(h_i')<w<\lambda$,
		then
		$	gcd(h_i+h_j, f_{p,\tau}(x))=1, \forall 0\leq i <j <n$.
		This indicates that any $4\times 4$ sub-matrix of $H$ is invertible over $\mathbb{R}_{p,\tau}$.
		Next, we focus on the remaining cases.
		We only need to determine if the following matrices are invertible over $\mathbb{R}_{p,\tau}$,
		\begin{equation}\label{eq:26--}
			\begin{pmatrix}
				1 & 1\\
				h_i^3 &h_j^3 
			\end{pmatrix},\quad 
			\begin{pmatrix}
				1 & 1& 1\\
				h_i &h_j &h_k\\
				h_i^3 &h_j^3 &h_k^3
			\end{pmatrix},\quad 
			\begin{pmatrix}
				1 & 1& 1\\
				h_i^2 &h_j^2 &h_k^2\\
				h_i^3 &h_j^3 &h_k^3
			\end{pmatrix},  
		\end{equation}
		where $0\leq i<j<k<n$.
		According to generalized Vandermonde determinants~\cite{kolokotronis2006lower}, the determinants of the above three matrices are respectively (let $h_{n-1}'=0$)
		\begin{equation}
			\begin{aligned}
				&h_i^3+h_j^3=(h_i+h_j)(1+x^\tau)^2 \cdot \left((h_i')^2+h_i'h_j'+(h_j')^2 \right.\\
				&\qquad \qquad   \left.+(h_i'+h_j')x^{w}+x^{2w}\right),\\
				&h_i+h_j+h_k=(1+x^\tau)(h_i'+h_j'+h_k'+x^{w}),\\
				&h_i h_j+h_i h_k+h_j h_k
				=(1+x^\tau)^2\\
			&\qquad \qquad  	\cdot (h_i' h_j'+h_i' h_k'+h_j' h_k'+x^{2w}),
			\end{aligned}
		\end{equation}
		where $0\leq i<j<k<n$.
		Since all $h_i', h_j',h_k'$ have degrees less than $w$, the above three values are not zero.
		Furthermore, due to $2w<\lambda$, they are all coprime with $f_{p,\tau}(x)$.
		Then all the matrices in \eqref{eq:26--} are invertible over $\mathbb{R}_{p,\tau}$.
		This completes the proof.
	\end{proof}

\end{construction}

\begin{remark}
	It is clear that all the codes in Construction~\ref{cons:1}, \ref{cons:2}, and \ref{cons:3} allow the total number of data columns to reach the exponential size with respect to the design parameter $p$.
	This is suitable for the needs of large-scale storage systems~\cite{lv2023new}.
	In addition, it is possible to construct the new codes using other matrices, such as Moore matrices~\cite{martinez2022general} and some matrices searched by computers. All proposed conditions in Section~\ref{sec:4.b} offer great flexibility in constructing the variant codes.
	
\end{remark}


\subsection{Fast Computations}\label{sec:05}

To begin with, one can know from coding theory that the product of any parity-check matrix and its corresponding codeword is zero~\cite{roth2006introduction}. Formally, $\mathbf{0}^{\mathrm{T}}=\widehat{H}\cdot \widehat{\mathbf{x}}^{\mathrm{T}}$, where $\mathbf{0}$ denotes a zero vector, $\widehat{H}$ denotes a binary parity-check matrix, and $\widehat{\mathbf{x}}$ denotes the corresponding codeword. 
It follows that $\widehat{H}\cdot \mathbf{x}^{\mathrm{T}}=\widehat{H}_e\cdot \mathbf{e}^{\mathrm{T}}$, where $\mathbf{x}$ denotes the codeword after all erased symbols are set to zero, $\mathbf{e}$ denotes the vector consisting of all erased symbols, and $\widehat{H}_e$ denotes the sub-matrix of $\widehat{H}$ corresponding to $\mathbf{e}$.
The above leads to the following common framework for encoding and decoding procedures~\cite{yu2020fast,yu2023reed}: (when encoding, all parity symbols can be regarded as erased symbols.)
\begin{enumerate}[Step 1.]
	\item Compute syndrome $\mathbf{s}^{\mathrm{T}}:=\widehat{H}\cdot \mathbf{x}^{\mathrm{T}}$.
	\item Solve linear equations $\mathbf{s}^{\mathrm{T}}=\widehat{H}_e\cdot \mathbf{e}^{\mathrm{T}}$.
\end{enumerate}

Note that in Step 2, $\mathbf{e}$ can be calculated by $\mathbf{e}^{\mathrm{T}}=\widehat{H}_e^{-1} \cdot \mathbf{s}^{\mathrm{T}}$.
In practice, each storage node holds a massive amount of data. Once all erased nodes (from power outages, downtime, etc.) are identified, the inverse of $\hat{H}_e$ needs to be computed only once to recover all data stored in erased nodes.
This results in the computational complexity of Step 2 being dominated by matrix-vector multiplication, which requires at most $c\cdot r^2(m-\tau)^2$ XOR,\footnote{In fact, this computational complexity can be reduced by scheduling algorithms for matrix-vector multiplication in the binary field, such as ``four Russians'' algorithm~\cite{chan2014speeding} or other heuristic algorithms in \cite{plank2012heuristics,huang2007optimizing}.} {where $c$ is a very large constant determined by the capacity of storage nodes.}
If $r,\tau$ are constants and $p = \Theta(\lg n )$, we have $\lim_{n \rightarrow \infty} \frac{c\cdot  r^2(m-\tau)^2}{c\cdot  (m-\tau)n}=0$, where $m=p\tau$.
This means that the asymptotic computational complexity of encoding/decoding is dominated by syndrome computation.
This subsection proposes fast syndrome computations for the constructed Vandermonde-based variant codes.

\subsubsection{Syndrome computation for Construction~\ref{cons:2}}\label{sec:v.a}
Here, $\widehat{H}=\mathcal{T}_{r, n,m}(H)$, then the syndrome computation is $\mathbf{s}^{\mathrm{T}} = \mathcal{T}_{r, n,m}(H) \cdot \mathbf{x}^{\mathrm{T}}$. Let $\mathbf{x}=(\mathbf{x}_0,..., \mathbf{x}_{n-1})$ and $\mathbf{s}=(\mathbf{s}_0,..., \mathbf{s}_{r-1})$ with each $\mathbf{x}_i\in \mathbb{F}_2^{m-\tau}, \mathbf{s}_i\in \mathbb{F}_2^{m-\tau}$.
For any $i \in [0, r)$, we have
\begin{equation}\label{eq:23-}
	\begin{aligned}
		\mathbf{s}_i^{\mathrm{T}} =& \sum_{j=0}^{2^{n_0}-1} \mathcal{A}_{\tau, \tau}(h_j^i) \cdot \mathbf{x}_j^{\mathrm{T}}\\
		=&\sum_{j=0}^{2^{n_0}-1} \mathcal{A}_{\tau, \tau}\left((h_j')^i\cdot (1+x^\tau)^i\right) \cdot \mathbf{x}_j^{\mathrm{T}}.
	\end{aligned}
\end{equation}
The following is dedicated to demonstrating that $\mathbf{s}$ can be calculated with the asymptotic complexity of $\lfloor \lg r \rfloor + 1$ XORs per data bit as $n_0$ increases.

We first focus on the auxiliary calculation, i.e.,
$
(\mathbf{s}_i^{*})^{\mathrm{T}}=\sum_{j=0}^{2^{n_0}-1} \mathcal{A}_{0, 0}\left((h_j')^i\cdot (1+x^\tau)^i\right) \cdot (\mathbf{x}_j^{*})^{\mathrm{T}},  
$
where $i\in [0, r)$, $\mathbf{s}_i^{*}\in \mathbb{F}_{2}^{m}$ and $\mathbf{x}_i^{*}=(\mathbf{x}_i, {0,0,\cdots,0} )\in \mathbb{F}_2^m$.
Obviously, for any $i\in [0,r)$, the first $m-\tau$ symbols in $\mathbf{s}_i^{*}$ exactly form $\mathbf{s}_i$.
Note that the auxiliary calculation can be converted into
\begin{equation}\label{eq:24}
	(\mathbf{s}_i^{*})^{\mathrm{T}}=\mathcal{A}_{0, 0}\left((1+x^\tau)^i\right)\cdot \sum_{j=0}^{2^{n_0}-1} \mathcal{A}_{0, 0}\left((h_j')^i \right) \cdot (\mathbf{x}_j^{*})^{\mathrm{T}},
\end{equation}
since $\mathcal{A}_{0, 0}$ is an isomorphic mapping.
%
%
%
%
In the above formula, the result of multiplying $\mathcal{A}_{0, 0}\left((h_j')^i \right)$ by $(\mathbf{x}_j^{*})^{\mathrm{T}}$ is in fact the reverse coefficient vector of the resultant polynomial from multiplying $(h_j')^i$ by
\begin{equation}\label{eq:25}
\mathbf{x}_j^{*}(x) := \mathbf{x}_j^{*}\cdot (x^{m-1},...,x,1)^{\mathrm{T}}.
\end{equation}
Hence, \eqref{eq:24} can be easily obtained after calculating the following polynomial multiplication
\begin{equation}\label{eq:26}
	P(i, \{\mathbf{x}_j^*(x)\}_{j=0}^{2^{n_0}-1}) :=   \sum_{j=0}^{2^{n_0}-1} (h_j')^i \cdot \mathbf{x}_j^{*}(x), \quad i\in [0, r).
\end{equation}
It can be seen from the setting of $\{h_j'\}_{j=0}^{2^{n_0}-1}$ that the calculation in \eqref{eq:26} is similar to the syndrome computation in \cite{yu2023reed}. The only difference is that the calculation is performed in the polynomial ring $\mathbb{R}$, while \cite{yu2023reed} is in a binary extension field.
%
Fast syndrome computation in \cite{yu2023reed} can be easily extended to the case of polynomial rings.
From \cite{yu2023reed}, we have the following lemma for computing \eqref{eq:26}.
\begin{lemma}
	Let $\mathbf{y}^{\mathrm{T}}=(\mathbf{y}_0,\mathbf{y}_1,...,\mathbf{y}_{2^{n_0}-1})^{\mathrm{T}}=R_{n_0} \cdot \left(\mathbf{x}_0^{*}(x),...,\mathbf{x}_{2^{n_0}-1}^{*}(x)\right)^{\mathrm{T}}$, where each $\mathbf{y}_0\in \mathbb{R}$, $R_{n_0}$ is a Reed-Muller matrix defined by $R_0=(1)$ and
	{
	\begin{equation}
			R_{i+1}=
		\begin{pmatrix}
			R_{i}&R_{i}\\
			\mathbf{0}_i&R_{i}
		\end{pmatrix},
	\end{equation}
	where $i\in \mathbb{N}$ and $\mathbf{0}_i$ denotes the $2^i\times 2^i$ all-zero matrix.}
	Then for any $i\in [0, r)$, 
\begin{equation}\label{eq:28}
	\begin{aligned}
		&P(i, \{\mathbf{x}_j^*(x)\}_{j=0}^{2^{n_0}-1})\\
		=&\begin{cases}
			\mathbf{y}_0, & \text{if}~b(i)=0,\\
			\sum_{j=0}^{n_0-1}x^{ij}\cdot \mathbf{y}_{2^j}+\sum_{0<j<2^{n_0}\atop 1< b(j)\leq b(i)}f(i,j)\cdot \mathbf{y}_{j},& \text{if}~b(i)\geq 1,
		\end{cases}
	\end{aligned}
\end{equation}
where $b(i)$ is the number of 1s in the binary representation of $i$, and $f(i,j)$ is a function that depends only on the indices $i$ and $j$.
In particular, when $b(i)=2$, each $f(i,j)$ in \eqref{eq:28} is a polynomial containing two terms.
\begin{proof}
	The proof can be easily obtained by analogy with that in \cite{yu2023reed}.
\end{proof}
\end{lemma}
From the above, the syndrome computation in \eqref{eq:23-} can be completed through the following steps (given $m,r$ and $n_0$):
\begin{enumerate}[Step 1.]
	\item From the input vector $(\mathbf{x}_0^{*}(x),...,\mathbf{x}_{2^{n_0}-1}^{*}(x))$, calculate all required $\mathbf{y}_i$ in \eqref{eq:28}.	
	\item From \eqref{eq:28}, calculate $\{P(i, \{\mathbf{x}_j^*(x)\}_{j=0}^{2^{n_0}-1})\}_{i=0}^{r-1}$.	
	\item Calculate $\{\mathbf{s}_i^*\}_{i=0}^{r-1}$ according to \eqref{eq:24}, and then extract $\{\mathbf{s}_i\}_{i=0}^{r-1}$.
\end{enumerate}

\begin{table*}[]
	\centering
	\caption{{Computational complexities of syndrome computations\\in the Vandermonde-based variant codes (\# of XORs per data bit)}}\label{tab:2}
	\begin{tabular}{cccccccc||c}
		\hline
		\multicolumn{9}{c}{Configurations }                                               \\ \hline \hline
		\multicolumn{1}{c||}{$p$}   & \multicolumn{3}{c|}{11 $(\lambda=10)$}                       & \multicolumn{3}{c|}{13$(\lambda=12)$}              & 17$(\lambda=8)$   &Theoretical \\ \cline{1-8}
		\multicolumn{1}{c||}{$n_0$ or $n_1$} & 8      & 9      & \multicolumn{1}{c|}{10}     & 8      & 9 & \multicolumn{1}{c|}{10} & 8   &Results   \\ \hline \hline
		\multicolumn{9}{c}{ V-ETBR$(n=2^{n_0}, r,m=p, H)$ }                                               \\ \hline \hline
		\multicolumn{1}{c||}{$r=3$} & 2.026 & 2.015 & \multicolumn{1}{c|}{2.008} & 2.027 & 2.015 & \multicolumn{1}{c|}{2.008}   &  2.028 & 2\\  
		\multicolumn{1}{c||}{$r=4$} & 3.112 & 3.070 & \multicolumn{1}{c|}{3.043} & 3.117 & 3.073  & \multicolumn{1}{c|}{3.045}   & 3.123  & 3\\
		\multicolumn{1}{c||}{$r=5$} & 3.145 & 3.088 & \multicolumn{1}{c|}{3.053} & 3.150 & 3.091  & \multicolumn{1}{c|}{3.055}   & 3.156 & 3\\
		\multicolumn{1}{c||}{$r=6$} & 3.376 & 3.234 & \multicolumn{1}{c|}{3.143} & 3.384 &3.240   & \multicolumn{1}{c|}{3.146}   & 3.395 & 3\\ 		
		\multicolumn{1}{c||}{$r=7$} & 3.607& 3.380 & \multicolumn{1}{c|}{3.232} & 3.619 &3.387   & \multicolumn{1}{c|}{3.237}   & 3.635 & 3\\ 				
		\multicolumn{1}{c||}{$r=8$} & 5.795& 5.223 & \multicolumn{1}{c|}{4.807} & 5.874 &5.283   & \multicolumn{1}{c|}{4.848}   & 5.995 & 4\\ \hline \hline
		\multicolumn{9}{c}{ V-ESIP$(n=2^{n_1}+1, r=4,m=p, H')$ }                                               \\ \hline \hline
		\multicolumn{1}{c||}{$r=4$} & 3.118&  3.073& \multicolumn{1}{c|}{3.044 } & 3.191 &  3.075  & \multicolumn{1}{c|}{3.046 }   & 3.126 & 3
		\\ \hline
	\end{tabular}
\end{table*}

In Step 1, many operations involving zeros can be eliminated, as each $\mathbf{x}_i^*$ is obtained by filling zeros with $\mathbf{x}_i$.
In Step 2, if $r < 8$, all involved multiplications can be calculated using at most one vector addition and one circular shift. This is due to the fact that each multiplication factor is a polynomial containing no more than two terms.
%
%
%
If $r \geq 8$, it is best to use matrix-vector multiplication for this operation
(the multiplication of two polynomials over $\mathbb{R}$ can be converted into multiplying a circulant matrix by a coefficient vector of a polynomial).
This is because $f(i,j)$ in \eqref{eq:28} contains too many terms that need to be summed.
In contrast, when implemented using matrix-vector multiplication, there exist general scheduling algorithms that can reduce the computational complexity.
In Step 3, the involved two operations can be merged into 
\begin{equation}
	\mathbf{s}_i^{\mathrm{T}}=\mathcal{A}_{\tau, 0}\left((1+x^\tau)^i\right)\cdot \sum_{j=0}^{2^{n_0}-1} \mathcal{A}_{0, 0}\left((h_j')^i \right) \cdot (\mathbf{x}_j^{*})^{\mathrm{T}},
\end{equation}
where $i\in [0,r)$.

In terms of complexity, Step 1 requires only a portion of the RM transform, and one can know from \cite{yu2023reed} that it produces XORs with the number of $(m-\tau) \cdot \left((\lfloor \lg r \rfloor + 1)n + \small{o}(n)\right)$~\cite{yu2023reed}, where little-o notation is used to describe an upper bound that cannot be tight.
Step 2 produces multiplications and additions that are both $\sum_{i=1}^{r-1} \sum_{t=1}^{b(i)} \binom{n_0}{t}-r+1$. 
When $r$ is a constant, it is not difficult to check that $\lim_{n_0\rightarrow \infty} \frac{\sum_{i=1}^{r-1} \sum_{t=1}^{b(i)} \binom{n_0}{t}}{2^{n_0}/n_0}=0$. Thus, the total number of XORs required for Step 2 is $m^2\cdot \small{o}(2^{n_0}/n_0)$.
Step 3 produces $r-1$ matrix-vector multiplications.
In summary, when $r$ and $\tau$ are constants and $n=2^{n_0}$ approaches infinity, the asymptotic complexity of the above syndrome computation is $\lfloor \lg r \rfloor + 1$ XORs per data bit. Note that $m=p\tau$ and $p=\Theta(n_0)$, where big-$\Theta$ notation is used to describe a bound within a constant factor.
For visualization, TABLE~\ref{tab:2} lists the computational complexities required for the proposed syndrome computation with different parameters.
It can be observed that the numerical results are close to the theoretical ones, especially when $n_0$ is large enough.
Indeed, the syndrome computation proposed in \cite{lv2023new}, which reaches an asymptotic complexity of two XORs per data bit,  is a special case of the above scheme at $r=3$.


\subsubsection{Syndrome computation for Construction~\ref{cons:3}}\label{sec:v.b}
Here, let $\mathbf{x}=(\mathbf{x}_0,...,\mathbf{x}_{n+3})$ of each $\mathbf{x}_i\in \mathbb{F}_2^{m-\tau}$ be a codeword, and $\mathbf{s}=(\mathbf{s}_0,..., \mathbf{s}_{3})$ of each $\mathbf{s}_i\in \mathbb{F}_2^{m-\tau}$ the corresponding syndrome. Note that in Construction~\ref{cons:3}, $n=2^{n_1}+1$ and the parity-check matrix $\mathcal{T}_{r,n,m}(H')$ is systematic.
For any $i\in [0,4)$, we have
\begin{equation}\label{eq:-34-}
	\begin{aligned}
		&\mathbf{s}_i^{\mathrm{T}} = \mathbf{x}_{2^{n_1}+i}^{\mathrm{T}}+\sum_{j=0}^{2^{n_1}-1} \mathcal{A}_{\tau, \tau}(h_j^i) \cdot \mathbf{x}_j^{\mathrm{T}}\\ =&\mathbf{x}_{2^{n_1}+i}^{\mathrm{T}}+  \sum_{j=0}^{2^{n_1}-1} \mathcal{A}_{\tau, 0}\left((h_j'+x^w)^i(1+x^\tau)^i\right) \cdot (\mathbf{x}_j^*)^{\mathrm{T}}  \\
		=&\mathbf{x}_{2^{n_1}+i}^{\mathrm{T}}+\mathcal{A}_{\tau, 0}\left((1+x^\tau)^i\right) \sum_{j=0}^{2^{n_1}-1} \mathcal{A}_{0, 0}\left((h_j'+x^w)^i\right) \cdot (\mathbf{x}_j^*)^{\mathrm{T}},
	\end{aligned}
\end{equation}
where each $\mathbf{x}_j^{*}=(\mathbf{x}_i,{0,0,...,0})\in \mathbb{F}_2^m$.
In the above formula, the result of multiplying $\mathcal{A}_{0, 0}\left((h_j'+x^w)^i \right)$ by $(\mathbf{x}_j^{*})^{\mathrm{T}}$ is in fact the reverse coefficient vector of the resultant polynomial from multiplying $(h_j'+x^w)^i$ by $\mathbf{x}_j^{*}(x) $, where $\mathbf{x}_j^{*}(x) $ is shown in \eqref{eq:25}.
Then, \eqref{eq:-34-} can be easily obtained after calculating the following polynomial multiplication
\begin{equation}
	Q(i, \{\mathbf{x}_j^*(x)\}_{j=0}^{2^{n_1}-1}) :=   \sum_{j=0}^{2^{n_1}-1} (h_j'+x^w)^i \cdot \mathbf{x}_j^{*}(x),   i\in [0, r).
\end{equation}
The above formula can be simplified as \eqref{eq:35}, which is shown at the bottom of this page.
This indicates that the syndrome computation can also be accelerated by \eqref{eq:28}.
From the above, the syndrome computation can be completed through the following steps:
\begin{enumerate}[Step 1.]
	\item From the input vector $(\mathbf{x}_0^{*}(x),...,\mathbf{x}_{2^{n_1}-1}^{*}(x))$, calculate all required $\mathbf{y}_i$ in \eqref{eq:35}.	
	\item Calculate $\{Q(i, \{\mathbf{x}_j^*\}_{j=0}^{2^{n_1}-1})\}_{i=0}^{3}$ according to \eqref{eq:35}.	
	\item Calculate \eqref{eq:-34-}.
\end{enumerate}
\begin{figure*}[b]
	\begin{equation}\label{eq:35}
		\begin{aligned}
			&	Q(i, \{\mathbf{x}_j^*(x)\}_{j=0}^{2^{n_1}-1})
			=
			\begin{cases}
				P(0, \{\mathbf{x}_j^*\}_{j=0}^{2^{n_1}-1}), &i=0,\\
				P(i, \{\mathbf{x}_j^*\}_{j=0}^{2^{n_1}-1})+ x^{iw} \cdot P(0, \{\mathbf{x}_j^*\}_{j=0}^{2^{n_1}-1})  , &i=1,2,\\
				{
					\begin{aligned}
						&P(3, \{\mathbf{x}_j^*\}_{j=0}^{2^{n_1}-1})+ x^{w} \cdot P(2, \{\mathbf{x}_j^*\}_{j=0}^{2^{n_1}-1})  
						+ x^{2w} \cdot P(1, \{\mathbf{x}_j^*\}_{j=0}^{2^{n_1}-1})\\&+ x^{3w} \cdot P(0, \{\mathbf{x}_j^*\}_{j=0}^{2^{n_1}-1}) 
					\end{aligned}
				}
				, &i=3.
			\end{cases}
		\end{aligned}
	\end{equation}
\end{figure*}

\begin{table*}[]
	\centering
	\caption{Asymptotic complexities of encoding/decoding when $r,\tau$ are constants and \\the total number of data columns approaches infinity (Per data bit).}\label{tab:3}
	\begin{tabular}{c|c|c|c|c||c}
		\hline
		MDS array codes                                          & Row size & Parity columns & Data columns & \# of XORs & Note                          \\ \hline \hline
		BR code~\cite{blaum1993new,subedi2013comprehensive}                                                & $p-1$    & $2\leq r<p$          & $p$                            & $r$                                                     & $p$ odd prime                 \\
		IP code~\cite{blaum1996mds,hou2018unified}                              & $p-1$    & $r\geq 2$            & $p+r$                        & $r$                                                     & $p$ odd prime                 \\
		Generalized RDP code~\cite{blaum2006family,hou2018unified}                                  & $p-1$    & $r\geq 2$            & $p+r-1$                        & $r$                                                     & $p$ odd prime                 \\
		Rabin-like code~\cite{hou2017new}                     & $p-1$    & $2\leq r<p$          & $p$                            & $2r$                                                    & $p$ odd prime                 \\ 
		Circulant Cauchy code~\cite{schindelhauer2013maximum} & $p-1$    & $2\leq r\leq p$            & $p+1$                          & $3r-2$                                                  & $2$ primitive element in $\mathbb{F}_p$ \\ 
		The Vandermonde-based V-ETBR code & $(p-1)\tau$    & $2\leq r< 2^{\lambda}$            & $2^{\lambda}$                          & $\lfloor \lg r \rfloor+ 1$                                                  & $p$ odd number\\  
		The Vandermonde-based V-ESIP code & $(p-1)\tau$    & $r=4$            & $2^{\lfloor \frac{\lambda-1}{2} \rfloor}+4$                          & 3                                                   &  $p$ odd number\\ 
		\hline
	\end{tabular}
\end{table*}

\begin{figure}[t]
	\centering
	\includegraphics[width=0.49\textwidth, height=0.23\textwidth]{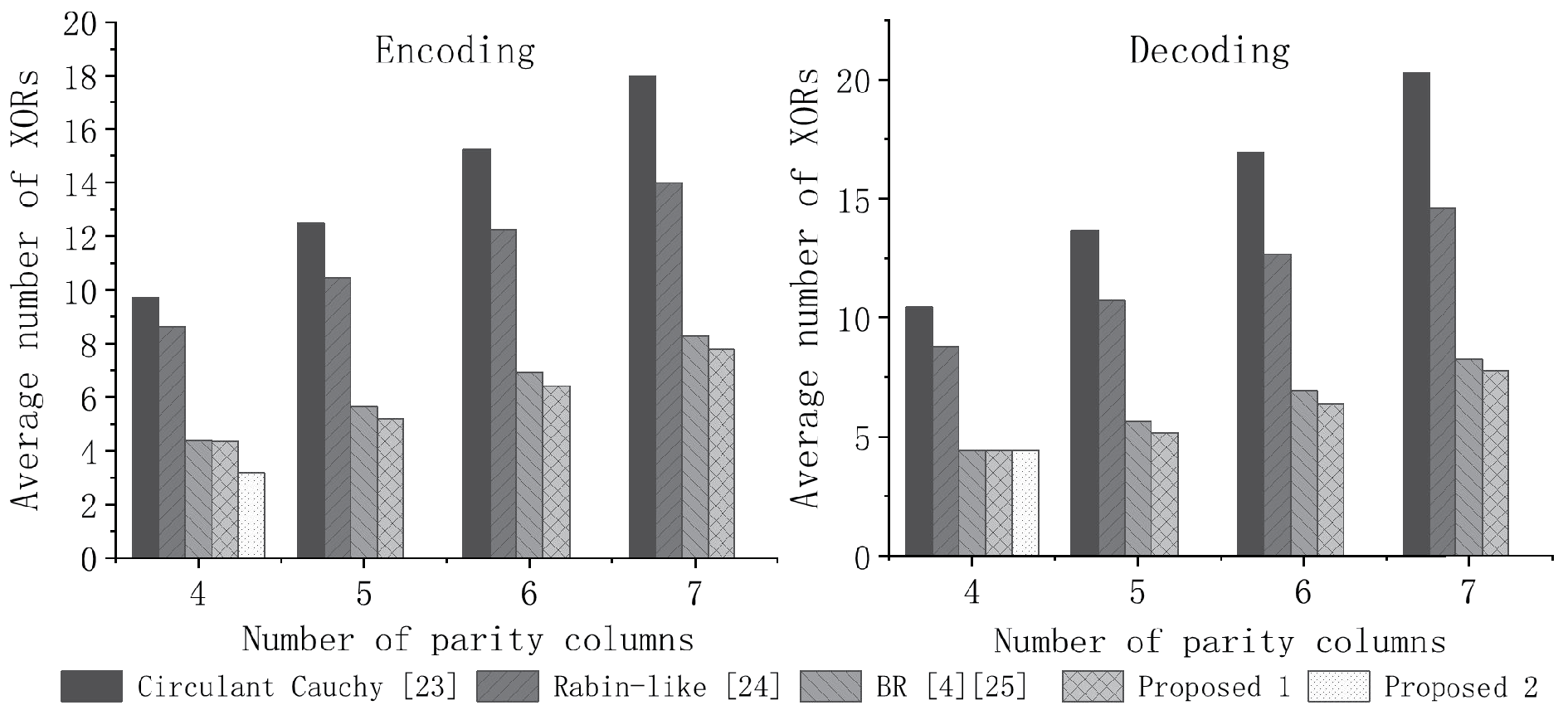}
	\caption{Computational complexities of different binary MDS array codes (when the total number of data columns is $127$).}
	\label{fig:3}
\end{figure}

\begin{figure}[t]
	\centering
	\includegraphics[width=0.49\textwidth, height=0.23\textwidth]{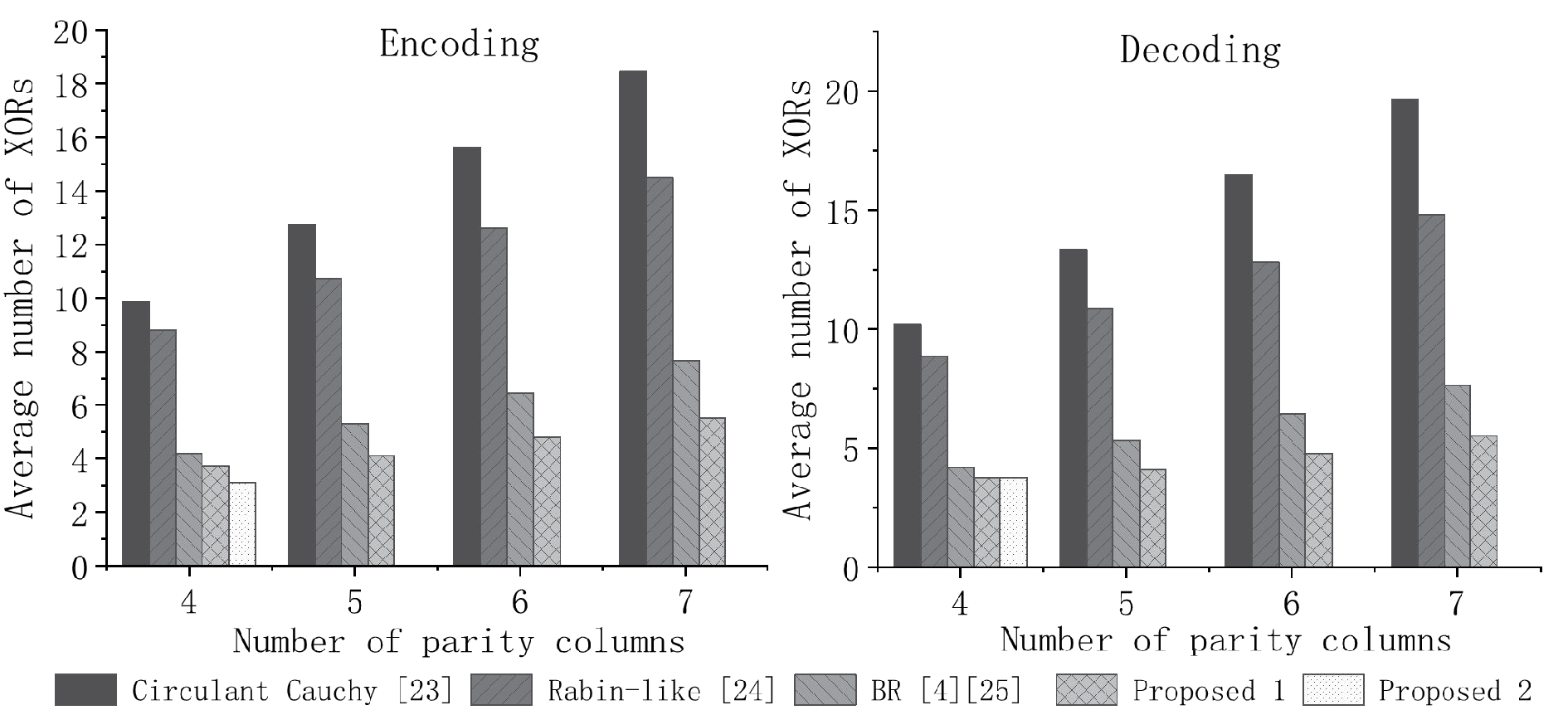}
	\caption{Computational complexities of different binary MDS array codes (when the total number of data columns is $251$).}
	\label{fig:4}
\end{figure}
In terms of complexity, Step 3 only requires a few vector additions and cyclic shifts.
When $r,\tau$ are constants and $n_0$ approaches infinity, the asymptotic complexity of the above is dominated by the first two steps, and it is obviously the same as that in Sec.~\ref{sec:v.a}, i.e., $\lfloor \lg r \rfloor + 1=3$ XORs per data bit.
TABLE~\ref{tab:2} also lists the computational complexities for this syndrome computation with different parameters.
Note that the total number of data columns at this time is $n+r-1$.

\subsection{Comparison}\label{sec:04}

TABLE~\ref{tab:3} lists the asymptotic complexities of different binary MDS array codes.
The fourth column shows the maximum number of data columns for each code, and the fifth column shows the asymptotic complexities of encoding and decoding, both of which are equal.
It can be observed that the constructed Vandermonde-based variant codes not only have a more flexible row size and design parameter $p$ but also have an exponentially growing total number of data columns with respect to $p$ and minimal asymptotic encoding/decoding complexity.

To better demonstrate the impact of asymptotic computational complexity in practice, Fig. \ref{fig:3} and \ref{fig:4} also show the average number of XORs required for different binary MDS array codes with the total number of data columns being 127 and 251, respectively.
Note that the average number of XORs is obtained by dividing the total number of XORs by the total number of bits in the data array,
and that ``Proposed 1" and ``Proposed 2" in Fig. \ref{fig:3} and \ref{fig:4} correspond to the Vandermonde-based V-ETBR and V-ESIP codes in TABLE~\ref{tab:3}, respectively.
In our setup, the parameters $p$ and $\tau$ of the variant codes are fixed to $p=11$ and $\tau=1$, while the parameter $p$ of the other codes are the same as the total number of data columns ($p=127$ in Fig.~\ref{fig:3}, $p=251$ in Fig. \ref{fig:4}).
Each code has a row size of $p-1$ in the data array.
This means that the row size in the data array of the variant codes is much smaller than that of other codes.
In other words, the proposed variant codes require significantly less capacity per node in storage systems.

Let the variant codes use ``Proposed 2" in the case of four parity columns and ``Proposed 1" in the other cases. Fig. \ref{fig:3} shows that the average improvements in encoding/decoding for the variant codes compared to the Circulant Cauchy code~\cite{schindelhauer2013maximum}, Rabin-like code~\cite{hou2017new}, and BR code~\cite{blaum1993new,subedi2013comprehensive}   are 60\%/61\%, 51\%/49\%, and 12\%/5\%, respectively. The average improvements in Fig. \ref{fig:4} are 69\%/69\%, 63\%/61\%, and 26\%/22\%, respectively.
With a fixed number of parity columns, the performance advantage of the variant codes in Fig. \ref{fig:4} is more obvious than that in Fig. \ref{fig:3}.

It is worth noting that	the practical performance of the variant codes constructed in this paper converges to the theoretical results when the number of data columns is much larger than that of parity columns.
When the total number of data columns is not large enough, the proposed syndrome computation does not dominate the overall computational complexity, causing the efficiency of binary matrix-vector multiplication to be crucial.
In our simulations, no additional scheduling algorithms for binary matrix-vector multiplication was used in the variant codes.
Thus, there is a great potential to further improve the performance of the variant codes, which is also one of our future work.

\section{Conclusion}\label{sec:5}

In this paper, we explore variant codes from codes over the polynomial ring $\mathbb{F}_2[x]/\langle \sum_{i=0}^{p-1}x^{i\tau}\rangle$, and then propose two new classes of binary array codes, termed V-ETBR and V-ESIP codes.
These variant codes are derived by mapping parity-check matrices over the polynomial ring to binary parity-check matrices.
We show that the well-known generalized RDP code is a special case of the variant codes.
To make this mapping a powerful tool in the construction of binary array codes, we explore in detail the connections between the variant codes and their counterparts over the polynomial ring, and provide conditions that make them binary MDS array codes.
Based on these conditions, some new binary MDS array codes are explicitly constructed based on Cauchy and Vandermonde matrices.
In addition, two fast syndrome computations for the constructed Vandermonde-based codes are proposed, both of which meet the lowest known asymptotic complexity among MDS codes~\cite{yu2023reed}.
Since the constructed codes have significantly more data columns than previous binary MDS array codes, the known lowest asymptotic computational complexity, and they are constructed from simpler binary parity-check matrices, they are attractive in practice.


\bibliographystyle{IEEEtran}
\bibliography{ref}

\end{document}